\documentclass[preprint,11pt]{elsarticle}
\usepackage{threeparttable}
\usepackage{fancyvrb}
\usepackage{lmodern}% http://ctan.org/pkg/lmodern
\usepackage{slantsc}% http://ctan.org/pkg/slantsc
\usepackage{bold-extra}
\usepackage[T1]{fontenc}
\usepackage{float}
\usepackage{soul}
\usepackage{color}

\def\mrg#1{{\color{black}{#1}}} % Michel
\def\ytl#1{{\color{black}{#1}}} % Yanting
\def\per#1{{\color{black}{#1}}} % Per
\def\gg#1{{\color{black} #1}} % Gediminas
\def\cyc#1{{\color{black} #1}} % Chongyang Chen

\usepackage{amssymb}
\usepackage{amsthm}
\usepackage{amsmath,amssymb}
\usepackage{bm}
\usepackage[normalem]{ulem} %CYC
\usepackage{hyperref}
\usepackage{multirow}
\usepackage{graphicx}
\usepackage{natbib}
\def\hhrule{\vspace{6pt}\hrule\vspace{6pt}}

\newcounter{bla}

\usepackage[dvipsnames]{xcolor}
\usepackage{subfig}
\usepackage{listings}
\usepackage{hyperref}
\makeatletter
\AtBeginDocument{%
  \let\c@figure\c@lstlisting
  
  \let\ftype@lstlisting\ftype@figure % give the floats the same precedence
}
\makeatother

\usepackage{tikz}
\usetikzlibrary{shapes}

\journal{Computer Physics Communications}

\begin{document}

\begin{frontmatter}



\title{{\sc Graspg} -- An extension to {\sc Grasp}2018 based on Configuration State Function Generators}


\author[a]{Ran Si}
\author[a]{\ytl{Yanting Li}}
\author[b]{Kai Wang}
\author[a]{Chongyang Chen \corref{author}}
\author[c]{Gediminas Gaigalas \corref{author}}
\author[d]{\mrg{Michel Godefroid}}
\author[e]{Per J\"onsson \corref{author}}

\cortext[author] {Corresponding author.\\\textit{E-mail address:} chychen@fudan.edu.cn, gediminas.gaigalas@tfai.vu.lt, per.jonsson@mau.se}
\address[a]{
Shanghai EBIT Lab, Key Laboratory of Nuclear Physics and Ion-Beam Application, Institute of Modern Physics, Department of Nuclear Science and Technology, Fudan University, Shanghai 200433, China}
\address[b]{ Department of Physics, Anhui Province Key Laboratory for Control and Applications of Optoelectronic Information Materials, Anhui Normal University, Wuhu 241000, Anhui, China}
\address[c]{
Institute of Theoretical Physics and Astronomy, Vilnius University, Saulėtekio av. 3, LT-10222 Vilnius, Lithuania}
\address[d]{Spectroscopy, Quantum Chemistry and Atmospheric Remote Sensing, Universit\'e libre de Bruxelles, Belgium}
\address[e]{
Department of Materials Science and Applied Mathematics, Malmö University, SE-20506 Malmö, Sweden}

\begin{abstract}
\noindent
The {\sc Graspg} program package is an extension of {\sc Grasp}2018  [Comput. Phys. Commun. 237 (2019) 184-187] based on configuration state function generators (CSFGs). The generators keep spin-angular integrations at a minimum and  reduce substantially the execution time and the memory requirements for large-scale multiconfiguration Dirac-Hartree-Fock (MCDHF) and relativistic configuration interaction (CI) atomic structure calculations.
The package includes the improvements reported in [Atoms 11 (2023) 12] in terms of redesigned and efficient constructions of direct- and exchange potentials, as well as Lagrange multipliers, and  additional parallelization of the diagonalization procedure.  
Tools have been developed for predicting configuration state functions (CSFs) that are unimportant and can be discarded for 
large MCDHF or CI calculations based on results from smaller calculations, thus providing efficient methods for {\em a priori} condensation.
The package provides a seamless interoperability with {\sc Grasp2018}. From extensive test runs and benchmarking, we have demonstrated
reductions in the execution time and disk file sizes with factors of 37 and 98, respectively, for MCDHF calculations based on large orbital sets compared to corresponding {\sc Grasp2018} calculations. For CI
calculations, reductions of the execution time with factors over 200 have been attained. 
With a sensible use of the new possibilities for {\em a priori} condensation, CI calculations with nominally hundreds of millions of CSFs can be handled. 
%\cyc{Also compared to t he corresponding {\sc Grasp2018} calculations, here computation loads (CPU time and disk-file sizes) are demonstrated to be reduced %by factors of above XXX or even more for the spin-angular integration calculation, thanks to the implementation of CSFG; The CPU time for MCDHF calculation %could be reduced by factors of XXX or even more, thanks to the improvements presented in [Atoms 11 (2023) 12].} 
\end{abstract}


\end{frontmatter}

% \linenumbers

% Computer program descriptions should contain the following
% PROGRAM SUMMARY.

\noindent
{\bf PROGRAM SUMMARY}\smallskip\\
  %Delete as appropriate.
\begin{small}
\noindent
{\em Program Title:} {\sc Graspg}                                           \smallskip\\
%{\em Program Files doi:} .  \\
{\em Licensing provisions: MIT License.}                                    \smallskip\\
{\em Programming language:} Fortran 95                        \smallskip\\
{\em Nature of problem:} Prediction of atomic energy levels using a multiconfiguration Dirac–Hartree–Fock approach. \smallskip\\
  %Describe the nature of the problem here. \\
{\em Solution method:} 
The computational method is the same as in {\sc Grasp2018} [1] except that configuration state function generators (CSFGs) have been introduced, a concept
that substantially reduces the execution time and memory requirements for large-scale calculations [2].  The method also relies
on redesigned and more efficient constructions of  \mrg{direct}
and exchange potentials, as well as Lagrange multipliers, along with
additional parallelization of the diagonalization procedure as detailed \mbox{in [3]}.
 \smallskip\\
  %Describe the method solution here. 
{\em Additional comments, including Restrictions and Unusual features :} 1. provides a seamless
interoperability with {\sc Grasp}2018,  2. options to limit the Breit interaction, 3. includes tools for predicting CSFs that are
unimportant and can be discarded for large MCDHF or CI calculations based
on the results from smaller calculations.\medskip\\
{\bf References}
\begin{enumerate}
\item 
C. Froese Fischer, G. Gaigalas, P. Jönsson, and J. Bieroń, Comput. Phys. Commun. 237 (2019) 184-187.
\item 
Y. Li, K. Wang, S. Ran {\em et al.} Comput. Phys. Commun. 283 (2023), 108562.
\item
Y. Li, J. Li, C. Song {\em et al.} Atoms 11 (2023) 12.
\end{enumerate}

% main text
\end{small}

\vspace{0.2cm}

\section{Introduction}
This paper describes the {\sc Graspg} program package. 
The package is an extension to {\sc Grasp2018}~\cite{GRASP2018} and allows atomic structure calculations to be performed faster 
and with the use of less resources in terms of disk space and memory.

The {\sc Grasp2018} program package implements the relativistic multiconfiguration method, and wave functions
of the targeted states are obtained as expansions over configuration state functions (CSFs) built on relativistic one-electron orbitals. 
In multiconfiguration Dirac-Hartree-Fock (MCDHF) calculations the radial parts of the one-electron orbitals are obtained by solving
coupled integro-differential equations, whereas the mixing coefficients of the CSFs are obtained by computing and diagonalizing the Hamiltonian matrix. The two steps are iterated until self-consistency has been obtained.  
In configuration interaction (CI) calculations the radial parts of the one-electron \mrg{orbitals} are available from previous MCDHF calculations, 
and only the mixing coefficients of the CSFs are determined by computing and diagonalizing the Hamiltonian matrix~\cite{IAN,Review,GRASPtheory}. An integral and time-consuming part of MCDHF and CI calculations is the computation of the Hamiltonian matrix elements.
In {\sc Grasp2018} the computation relies on spin-angular integration in coupled tensorial form~\cite{Gaigalas_1996,GG1}, which gives a resolution into spin-angular coefficients that are multiplied with radial integrals
and effective interaction strengths.

%Accurate calculations,
%accounting for valence-valence, core-valence, and core-core electron correlation, often rely on large %orbital sets and massive CSF expansions obtained from single- and double (SD) replacements from %multireferences (MR) consisting of the most important configurations~\cite{Review}.
%Analyzing the CPU time used for these type of calculations, it is seen that most time is spent on constructing the Hamiltonian matrix~\cite{GRASPCPU}.

The {\sc Graspg} program package is based on configuration state function generators (CSFGs). A \mrg{single} CSFG consists of a generating CSF with the principal quantum numbers
of symmetry ordered correlation orbitals at their highest, as allowed by the orbital set, and an orbital de-excitation rule. 
The generating CSF and the de-excitation rule act as a generator for a set of CSFs. 
%The computation of the Hamiltonian matrix element between two CSFs relies on spin-angular integration~\cite{ }, which gives a resolution into spin-angular coefficients that are multiplied with radial integrals
%and effective interaction strengths and then accumulated. 
With the use of CSFGs, spin-angular integrations need to be done only between the generating \mrg{CSFs}. Matrix elements between all the generated CSFs then follow directly
by applying the de-excitation rules to the radial integrals and the  effective interaction strengths.
This procedure is very efficient and has shown to reduce the execution time for computing the Hamiltonian matrix for CSF expansions built on large orbital sets with factors up to 68, see ~\cite{CSFG} and tables therein. {\sc Graspg} also includes the improvements reported in \cite{GRASPCPU}, i.e., redesigned 
and more efficient constructions of \mrg{direct} and exchange potentials, as well as Lagrange multipliers, along with additional parallelization of the diagonalization procedure, further reducing the execution time for both \mrg{MCDHF} and CI calculations. 

%In addition, it substantially reduces the amount of data needed to construct the differential equations in MCDHF calculations.  

Just like {\sc Grasp2018}, the new {\sc Graspg} program package 
 is file driven~\cite{GRASP2018_Man}. Internally, it uses files with generating CSFs, but 
 produces mixing coefficients relative to the full list of CSFs, obtained by applying the de-excitation rules to the generating CSFs,
 in a file format that is compatible with the one
 used by the programs and tools in {\sc Grasp2018}. Thus, it provides a seamless interoperability
 with {\sc Grasp2018}.
 
 The paper is organized as follows: first, we describe multiconfiguration methods as implemented in the  {\sc Grasp2018} package. Then we introduce the concept of CSFGs and show how it reduces the need for spin-angular integrations, thus not only speeding up the computation of the Hamiltonian matrix, but also reducing the amount of data
 to be stored in primary memory or on disk in order to construct the effective potentials of the coupled integro-differential equations for the radial orbitals in the self-consistent calculations. This is followed by a brief discussion of the redesigned constructions of the \mrg{direct}
and exchange potentials and  the Lagrange multipliers.
 After the presentation of the methodology, we provide the details on how to compile the  {\sc Graspg} package and merge it into an existing {\sc Grasp2018} installation. The structure of the {\sc Graspg}  package in terms of new libraries and application programs is presented. 
 This is followed by an account of the new input and output data files and the file flow. Finally, we provide a number of scripts and test runs to validate 
 the program operation. The test runs also form the basis for timing and disk use studies. All test runs were performed on a computer cluster with the following configuration: AMD EPYC 7542 32-Core Processor, CPU MHz 1500, 1 T\cyc{B} memory, type DDR4, with configured memory speed of 3200 MT/s, and a
 14 TB ATA disk.
 \section{Multiconfiguration methods and spin-angular integration}
\subsection{Multiconfiguration methods}
In the multiconfiguration method, as implemented in {\sc Grasp2018}, the wave function of an atomic state $\Gamma JM_J$
 is approximated by 
an atomic state function (ASF), which is an expansion over CSFs
\begin{equation}
\label{eq:jj_ASF}
\Psi(\Gamma JM_J)  = \sum_{i=1}^{M}
%c_{\gamma_{iJ}} 
c_{\gamma_{i}}
\Phi(\gamma_{i} JM_J).
\end{equation}
Here $\gamma$ specify the orbital \mrg{occupancies} and spin-angular coupling tree quantum numbers of the CSFs.
The CSFs are antisymmetrized and symmetry-adapted
all-electron functions, recursively constructed from products of relativistic one-electron
orbitals 
\begin{equation}
\label{eq:R_orbitals}
\psi_{nlsjm} (r, \theta, \varphi) = \frac{1}{r}
 \left( \begin{array}{c}
P_{nlj}(r) \; {\Omega}_{l s j m}(\theta, \varphi) \\
\mbox{i} \; Q_{nlj}(r) \; {\Omega}_{\tilde{l} s j m}(\theta, \varphi)  \end{array} \right),
\end{equation}
where $P_{nlj}(r)$ and $Q_{nlj}(r)$
 are the radial functions  
and ${\Omega}_{l s j m}(\theta, \varphi)$
are two-component spherical spinors, see \cite{IAN,Review,GRASPtheory} for details.

Multiconfiguration methods rely on the variational principle \mrg{\cite{FroGod:Handbook}}.
Requiring the energy computed from the multiconfiguration expansion to be
stationary with respect to perturbations in the mixing coefficients leads to a matrix
eigenvalue problem
\begin{equation}
({\bm H} - E {\bm I}){\bm c} = {\bm 0},
\end{equation}
where ${\bm c} = (c_1,c_2,\ldots,c_M)^T$ is the vector of mixing coefficients and ${\bm H}$ is the Hamiltonian matrix with elements 
$H_{ij} = \langle \Phi(\gamma_{i} J) \| {\cal H}_{\mbox{\scriptsize DC}} \| \Phi(\gamma_{j} J) \rangle $ of the Dirac-Coulomb (DC) Hamiltonian.
The stationary condition of the energy with respect to variations in the radial
functions, in turn, leads to a system of coupled integro-differential equations subject to
boundary conditions at the origin and the infinity. Calculations where both the radial functions 
and the mixings are determined are referred to as 
MCDHF calculations. Calculations for which the radial functions are known, and only the mixing coefficients are determined, are referred to as CI calculations. For the latter, the Breit interaction as well as the leading QED corrections can be added to the Dirac-Coulomb Hamiltonian, details can be found in \cite{IAN,Review,GRASPtheory}.
\subsection{Computation of reduced matrix elements}\label{ME}
The computation of matrix elements between CSFs, as needed for both MCDHF and CI calculations, rests on spin-angular integration~\cite{GRASPGG}, which gives spin-angular coefficients that are multiplied with radial integrals
and effective interaction strengths and then accumulated \cite{GG1,RED_TENSOR}. For the reduced matrix elements of the 
Dirac-Coulomb-Breit (DCB) Hamiltonian  we have
\cite{IAN,GRASPtheory}
%\begin{equation}
\begin{eqnarray}
\label{eq:MAT_ELT}
\hspace*{-2.0cm}
H_{ij} & = & \langle \Phi(\gamma_{i} J )\| {\cal H}_{\mbox{\scriptsize DCB}} \|   \Phi(     \gamma_{j} J )\rangle \nonumber  \\ [0.2cm]
& = & \underbrace{\sum_{ab} t_{ab}^{ij} I(a,b)  + \sum_{abcd;k} v_{abcd;k}^{ij} R^{k}(ab,cd)}_{\mbox{\scriptsize Dirac-Coulomb}} + 
\underbrace{\sum_{abcd;k} w_{abcd;k}^{ij} X^{k}(ab,cd),}_{\mbox{\scriptsize Breit}} 
\end{eqnarray}
%\end{equation}
where the sums are over the one-electron orbitals used for the constructions of the interacting CSFs.
The quantities $t_{ab}^{ij}$, $v_{abcd;k}^{ij}$ and $w_{abcd;k}^{ij}$ are the spin-angular coefficients.
The radial one-electron Dirac integrals are given by      
\begin{eqnarray}
\label{eq:I(a,b)}
%\hspace*{-1.5cm}
I(a,b)  \nonumber\\
&\hspace*{-1.0cm}=&\hspace*{-0.5cm}  \delta_{\kappa_a \kappa_b} \int_0^{\infty}
\left\{ P_{n_a\kappa_a}(r) V_{\mbox{\scriptsize nuc}}(r) P_{n_b\kappa_b}(r) 
-c P_{n_a\kappa_a}(r)\left(\frac{d\;}{dr}-\frac{\kappa}{r}\right) Q_{n_b\kappa_b}(r) \right. \nonumber\\  [0.1cm]
 &\hspace*{-1.0cm}+&\hspace*{-0.5cm} c\; Q_{n_a\kappa_a}(r) \left(\frac{d\;}{dr} +\frac{\kappa}{r}\right)P_{n_b\kappa_b}(r) \nonumber\\  [0.1cm]
 &\hspace*{-1.0cm}+&\hspace*{-0.5cm} \bigg.  Q_{n_a\kappa_a}(r)\Big(V_{\mbox{\scriptsize nuc}}\left(r\right)
-2c^2\Big)Q_{n_b\kappa_b}(r) \bigg\}dr .
\end{eqnarray}
The radial two-electron Coulomb integrals, the so-called Slater integrals, are given by
%\begin{eqnarray}
%\label{eq:rel-rk}
%R^k(ab,cd) = & &
%\int_0^{\infty}
%\left[ 
%P_{n_a\kappa_a}(r)P_{n_c\kappa_c}(r) \right.  \nonumber\\
%& & \left.+Q_{n_a\kappa_a}(r)Q_{n_c\kappa_c}(r)
%\right] \frac{1}{r} Y^{k}(bd;r)dr.
%\end{eqnarray}
\begin{eqnarray}
\label{eq:rel-rk}
R^k(ab,cd)  = 
\int_0^{\infty}
\Big[ 
P_{n_a\kappa_a}(r)P_{n_c\kappa_c}(r) +Q_{n_a\kappa_a}(r)Q_{n_c\kappa_c}(r)
\Big] \frac{1}{r} Y^{k}(bd;r)dr.
\end{eqnarray}
with $Y^k$ defined by 
%\begin{eqnarray}
%\label{eq:rel-yk}
% Y^{k}(ab;r) = & & r 
%\int_0^{\infty}
%\frac{r^k_{<}}{r^{k+1}_{>}}
%\left[ 
%P_{n_a\kappa_a}(s)P_{n_b\kappa_b}(s) \right.  \nonumber\\
%& & \left.
%+Q_{n_a\kappa_a}(s)Q_{n_b\kappa_b}(s)
%\right] ds.
%\end{eqnarray}
\begin{eqnarray}
\label{eq:rel-yk}
\hspace*{-0.6cm}
 Y^{k}(ab;r) =  r 
\int_0^{\infty}
\frac{r^k_{<}}{r^{k+1}_{>}}
\Big[ 
P_{n_a\kappa_a}(s)P_{n_b\kappa_b}(s) 
+Q_{n_a\kappa_a}(s)Q_{n_b\kappa_b}(s)
\Big] ds.
\end{eqnarray}
In the $Y^k$ integral $r_<$ and $r_>$ are the smaller and larger of $r$ and $s$, respectively. 
The Breit interaction is complex. In eq. (\ref{eq:MAT_ELT}) above,  $X^{k}(ab,cd)$ is the effective interaction strength given by
%\begin{equation}
\begin{eqnarray}
\label{eq:eff_int_strength}
%\hspace*{-1.5cm}
X^{k}(ab,cd) 
\nonumber\\  [-0.2cm]
& \hspace*{-1.8cm} = & \hspace*{-1.0cm} (-1)^k \langle j_a \| {\bm C}^{(k)} \| j_c \rangle \langle j_b \| {\bm C}^{(k)} \| j_d \rangle \sum_{\nu = k-1}^{k+1}
\Pi'(\kappa_a,\kappa_c,\nu) \Pi'(\kappa_b,\kappa_d,\nu)
   \nonumber\\  [-0.2cm]
&\hspace*{-1.8cm}& \hspace*{-1.0cm}\times \sum_{\mu = 1}^8 s_{\mu}^{\nu k}(a,b,c,d)S^{\nu}_{\mu}(a,b,c,d).
\end{eqnarray}
%\end{equation}
Here 
\begin{equation}
\langle j \| {\bm C}^{(k)} \| j' \rangle = (-1)^{j + 1/2} [j,j']^{1/2} 
\left(
\begin{array}{ccc}
j & k & j' \\
\frac{1}{2} & 0 & -\frac{1}{2} \\
\end{array}
\right)
\end{equation} 
and
\begin{equation}
\Pi'(\kappa,\kappa',\nu) =
\left\{
\begin{array}{ll}
1 & \mbox{if~} l + l' + \nu \mbox{~~is odd} \\
0 & \mbox{if~} l + l' + \nu \mbox{~~is even}. \\
\end{array}
\right.
\end{equation}
Further, $s_{\mu}^{\nu k}(a,b,c,d)$ are coefficients expressed in terms of the kappa quantum numbers and   
$S^{\nu}_{\mu}(a,b,c,d)$ radial double integrals. Details, along with definitions of the integrals, are given in~\cite{BreitGrant}.
\subsection{CI and Brillouin–Wigner perturbation theory}\label{sec:PT}
For accurate calculations, large numbers of CSFs are required, leading to large matrices. To handle these large matrices, the CSFs can be divided into two spaces. The first space,  referred to as the zero-order space, $P$, with $m$ elements ($m \ll M)$, contains CSFs that account for the major parts of the wave functions. 
The second space, referred to as the first-order space, $Q$, with $M-m$ elements, contains CSFs that represent minor corrections. 
Allowing interactions only between CSFs in $P$, interactions between CSFs in $P$ and $Q$, and diagonal interactions between CSFs in $Q$ 
gives a matrix
\begin{equation}
\left(
\begin{array}{cc}
H^{(PP)} & H^{(PQ)} \\
H^{(QP)} & H^{(QQ)} \\
\end{array}
\right),
\end{equation}
where $H^{(QQ)} = \delta_{ij}E^i$. The restriction to diagonal matrix elements in the large $Q$ space,
results in a huge reduction in the total number of matrix elements and the corresponding execution time. 
The above form of the matrix, yields energies that are similar to the ones obtained by applying second-order perturbation theory (PT) corrections to the energies of the smaller $m \times m$
matrix. The method is therefore sometimes referred to \mrg{as} CI combined with second-order Brillouin–Wigner perturbation theory \cite{Stefan, GPT}.
%\subsection{MCDHF calculations}\label{MCDHFcalc}
\section{Multiconfiguration calculations using {\sc Grasp}2018}\label{CIcalc}
\subsection{The CSF file format in {\sc Grasp}2018}
The starting point for both MCDHF and CI calculations is a list of CSFs. In {\sc Grasp}2018 the list of CSFs is generated by the {\tt rcsfgenerate} program by allowing excitations (single (S), double (D), triple (T), quadrupole (Q) etc.) according to some rule from a multireference  (MR) to a set of orbitals, see \cite{GRASP2018_Man} section 4. The CSFs are saved in a file
that starts with a specification of the closed core subshells, followed by the orbitals used to build the CSFs (the peel subshells). The above 
specifications 
are followed by the list of CSFs, where each CSF is written on three lines giving the \mrg{orbital} occupations and spin-angular couplings, see \cite{GRASP2018_Man} sections 4 and 5. As an example,  allowing SDTQ excitations from the $1s^22s2p$ reference configuration to the orbital set
\begin{equation}
\Big\{1s,2s,2p\mbox{-},2p,3s,3p\mbox{-},3p,3d\mbox{-},3d\Big\}
\end{equation}
along with
SD excitations to the orbitals set 
\begin{equation}
\Big\{4s,4p\mbox{-},4p,4d\mbox{-},4d,\ldots,7s,7p\mbox{-},7p,7d\mbox{-},7d\Big\} .
\end{equation}
we obtain the {\sc Grasp2018} file in table \ref{tab:CSF1} for the $J=0$ odd parity block. \mrg{This} file contains 1400 CSFs. Although the user may specify the orbital order, the default is by increasing $n$ and $l$, that is $1s,2s,2p\mbox{-},2p,3s,3p\mbox{-},3p,3d\mbox{-},3d$ etc. This order is referred to as the labeling order.  
\begin{table}
\caption{File with  CSFs  for the $J=0$ odd parity block obtained by the {\tt rcsfgenerate} program. The orbitals under the peel subshells heading are in labeling order, i.e., ordered by increasing $n$ and $l$.}\vspace{-3 mm}
\label{tab:CSF1}
\hhrule
\footnotesize{
\begin{verbatim}
Core subshells:

Peel subshells:
  1s   2s   2p-  2p   3s   3p-  3p   3d-  3d   4s   4p-  4p   4d-  
  4d   5s   5p-  5p   5d-  5d   6s   6p-  6p   6d-  6d   7s   7p- 
  7p   7d-  7d
CSF(s):
  1s ( 2)  2s ( 1)  2p-( 1)
               1/2      1/2
                           0-
  1s ( 2)  2s ( 1)  3p-( 1)
               1/2      1/2
                           0-
  1s ( 2)  2s ( 1)  4p-( 1)
               1/2      1/2
                           0-


            ...
            

  3p-( 1)  3d-( 2)  3d ( 1)
      1/2        2      5/2
                  5/2      0-
\end{verbatim}}
\hhrule
\end{table}
\subsection{Spin-angular integrations based on lists of CSFs}\label{sec:SACSF}
Denote the number of CSFs in the list by $M$.
Spin-angular integrations based on lists of CSFs are performed by nested double loops, where
the outer loop variable $i$ runs from 1 to $M$ and the inner loop variable $j$ runs from $i$ to  $M$.
For each reduced matrix element $H_{ij} = \langle \Phi(\gamma_{i} J )\| {\cal H} \|   \Phi(     \gamma_{j} J )\rangle$, where $ {\cal H}$ is the Dirac-Coulomb or the Dirac-Coulomb-Breit
Hamiltonian, there are  calls to the spin-angular routines of the  {\tt librang90} library  \cite{GRASPGG}. 
These routines
resolve the matrix
element into spin-angular coefficients and radial integrals. Consider, as an example, a list of four CSFs  with the orbitals in labeling order as shown in table \ref{tab:CSF2}.

\begin{table}[H]
\caption{List of CSFs with orbitals in labeling order. Spin-angular integration is performed for all reduced matrix elements  $H_{ij}$ such that $i \ge j$.}\vspace{-3 mm}
\label{tab:CSF2}
\hhrule
\footnotesize{
\begin{verbatim}
Core subshells:

Peel subshells:
  1s   2s   2p-  3s   3p-
CSF(s):
  1s ( 2)  2s ( 1)  2p-( 1)
               1/2      1/2
                           0-
  1s ( 2)  2s ( 1)  3p-( 1)
               1/2      1/2
                           0-
  1s ( 2)  2p-( 1)  3s ( 1)
               1/2      1/2
                           0-
  1s ( 2)  3s ( 1)  3p-( 1)
               1/2      1/2
                           0-
\end{verbatim}}
\hhrule
\end{table}
\noindent
Calls to the spin-angular routines resolves the Dirac-Coulomb matrix elements into spin-angular coefficients and pointers to radial one- and two-electron integrals, as shown in  table
\ref{tab:spin_ang}, where the permutation symmetries of $R^k$ integrals have been utilized to keep the number of integrals as small as possible. When writing angular data to file, the 
pointers to the integrals are stored as single integers in packed form. 

%\cyc{In specific, the pointers are sorted according to the indices \emph{abcd} in Eq.(\ref{eq:rel-rk}), and written into files named by mcpXXX.$32+k$ using the associated tensor rank $k$. mcp.30 stores the sparse structure of H matrix. mcp.31 stores the one-body spin-angular coefficients contributing to the effective potentials. \sout{In specific, the spin-angular coefficients are calculated by {\tt rangular} or {\tt rangular\_mpi}, and are stored in files mcpXXX.31, and mcpXXX.32,  and written into ... Should we add some statements about the structures of the mcp files?.... the pointers sorted, whereas the sorting procedure is removed in {\tt rangular\_csfg}, resulting also significant time reduction for large-scale calculation, compare the reduction by CSFG implementation. Some counterparts  for {\tt rangular\_csfg\_mpi} would be added in the corresponding section, where only three files mcp.30, mcp31, and mcp32 are output. Or, put the description for differences between {\tt rangular} and {\tt rangular\_csfg} in section \ref{sec:saCSFG} or \ref{sec:mcdhfCSFG}? }}
\clearpage

\begin{table}[h]
\caption{List of Dirac-Coulomb matrix elements between the CSFs in table \ref{tab:CSF2} resolved into spin-angular coefficients and pointers to radial one- and two-electron integrals.}
\label{tab:spin_ang}
\hhrule
\footnotesize{
\begin{verbatim}
    <1|H|1>
    2.000000000 I(1s ,1s )
    1.000000000 I(2s ,2s )
    1.000000000 I(2p-,2p-)
    1.000000000 R0(2s 2p-,2s 2p-)
   -0.333333333 R1(2s 2s ,2p-2p-)
    1.000000000 R0(1s 1s ,1s 1s )
    2.000000000 R0(1s 2s ,1s 2s )
   -1.000000000 R0(1s 1s ,2s 2s )
    2.000000000 R0(1s 2p-,1s 2p-)
   -0.333333333 R1(1s 1s ,2p-2p-)

    <2|H|1>
    1.000000000 I(2p-,3p-)
    1.000000000 R0(2s 2p-,2s 3p-)
   -0.333333333 R1(2s 2s ,3p-2p-)
    2.000000000 R0(1s 2p-,1s 3p-)
   -0.333333333 R1(1s 1s ,3p-2p-)

   .......
    
\end{verbatim}}
\hhrule
\end{table}

\subsection{MCDHF calculations in {\sc Grasp}2018}\label{sec:mcp}
MCDHF calculations are based on the Dirac-Coulomb Hamiltonian.
Before an MCDHF calculation can be performed, all the matrix elements between the CSFs have to be resolved into spin-angular coefficients and pointers to radial one- and two-electron integrals. This is done by the
{\tt rangular} or {\tt rangular\_mpi} programs, which both call the routines of the {\tt librang90}  library.
The spin-angular coefficients together with pointers to the corresponding one- and two-electron integrals are written to the unformatted {\tt mcp.30}, {\tt mcp.31}, {\tt mcp.32} ({\tt mcp.XXX})  files, which, for MPI calculations, are distributed over the directories specified by the {\tt disks} file, see \cite{GRASP2018_Man} \mbox{section 2}. At the next stage, the {\tt rmcdhf} or {\tt rmcdhf\_mpi} programs load the list of CSFs and the initial estimates of the radial orbitals, which are to be determined in the self-consistent procedure, into arrays. The programs then read the spin-angular coefficients and the pointers to the radial integrals from the {\tt mcp.XXX} files, compute the radial one- and two-electron integrals and construct the Hamiltonian matrix, which is diagonalized by a call to the Davidson eigenvalue solver \cite{Davidson} to give the mixing coefficients of the CSFs for the targeted states.
The programs again read the spin-angular coefficients and integral pointers from the {\tt mcp.XXX} files to construct the 
\mrg{direct} and exchange potentials of the coupled integro-differential equations for the radial orbitals, the solutions of which give the updated orbitals, see \cite{GRASPtheory} section 2.7 and \cite{GRASPCPU} section 3.2.
The above procedure is repeated until the energies and radial orbitals have converged within some predefined tolerance.

The repeated reading from file during the self-consistent procedure is time-consuming, though the modern server could well buffer the content. Provided the spin-angular data fit into the available memory, the memory-versions 
{\tt rmcdhf\_mem} and {\tt rmcdhf\_mem\_mpi} of the above programs load the contents of all the {\tt mcp.XXX} files into arrays at the program start, with a resulting reduction of the user time with factors up to 2, see \cite{GRASPCPU} table 2. 

\subsection{CI calculations in {\sc Grasp}2018}
When performing a CI calculation in {\sc Grasp2018}, the {\tt rci} or {\tt rci\_mpi} programs 
first evaluate the needed radial integrals and
effective interaction strengths and save them in memory. They then go on to compute the Hamiltonian matrix.
For each matrix element,
the programs perform the spin-angular integration by calls to routines of the {\tt librang90}  library \cite{GRASPGG}, multiplies the spin-angular coefficients with the radial integrals and effective interaction 
strengths and accumulates the result. The matrix elements are saved in the {\tt rci.res} files, which, for MPI calculations, are distributed over the directories specified by the {\tt disks} file.
The computation of the Hamiltonian matrix is followed  by a call to the Davidson eigenvalue solver \cite{Davidson}
 to give the mixing coefficients of the CSFs for the targeted states. Depending on the available memory per process, the 
 Hamiltonian matrix is loaded into memory or kept on disk during the diagonalization.
For large CSF expansions, the execution time for the CI calculations 
is dominated by the time for the computation of the Hamiltonian matrix \cite{GRASPCPU}. 
%To cut the time for the latter, 
%we note that spin-angular integration is independent of the principal quantum numbers of the orbitals and that this is something that can be exploited.
%Additionally, we note that the time for evaluating the Breit interaction can be reduced by posing limitations on the spin-angular quantum numbers of the participating orbitals, reflecting the fact that the Breit corrections to individual orbital energies drop rapidly with both $n$ and $l$, see for example ~\cite{Breitorbital}.
\section{Multiconfiguration calculations using {\sc Graspg}}
\subsection{Configuration state function generators}
To take advantage of the fact that spin-angular integration~\cite{Gaigalas_1996,GG1} is independent of the principal quantum numbers, and at the same time retain the possibility to label the states in accordance with the usual conventions in atomic spectroscopy, both the set of one-electron orbitals and the CSF space have to be rearranged. 
In our method, we divide the CSF space in what we call a labeling space and a correlation space. The CSFs in the labeling space
are obtained allowing excitations (SDTQ etc.) according to some general rule  
from an MR to a set of highly occupied orbitals in labeling order, e.g.,
\begin{equation}\label{eq:label}
\Big\{
1s,2s,2p\mbox{-},2p,3s,3p\mbox{-},3p,3d\mbox{-},3d
\Big\}.
\end{equation}
The CSFs in the labeling space account for major correlation effects due to close degeneracies and long-range rearrangements. Due to the fact that these
CSFs are built from orbitals in labeling order they can, after a transformation from $jj$- to $LSJ$-coupling as described by Gaigalas \cite{GG4b,GG4,GG5,ATOMSJJLSJ}, be used to label the atomic states in accordance with the conventions used, e.g., by NIST \cite{NIST}.
The CSFs in the correlation space are 
obtained by SD excitations from the MR to a set of symmetry ordered correlation orbitals. 
As a concrete example of the latter, we have
%begin{equation}
%\{4s,5s,6s,7s,4p\mbox{-},5p\mbox{-},6p\mbox{-},7p\mbox{-},4p,5p,6p,7p,4d\mbox{-},5d\mbox{-},6d\mbox{-},7d\mbox{-},
%4d,5d,6d,7d\}.
%\end{equation}
\begin{equation}\label{eq:sym}
\Big\{
\underbrace{4s,5s,6s,7s}_{s} ,
\underbrace{4p\mbox{-},5p\mbox{-},6p\mbox{-},7p\mbox{-}}_{p\mbox{-}} ,
\underbrace{4p,5p,6p,7p}_{p} ,
\underbrace{4d\mbox{-},5d\mbox{-},6d\mbox{-},7d\mbox{-}}_{d\mbox{-}} ,
\underbrace{4d,5d,6d,7d}_{d} 
\Big\}.
\end{equation}
The CSFs in the correlation space typically account for short range interactions and dynamical correlation \cite{Review}. 

The division of the orbitals into a set of labeling ordered orbitals and a set of symmetry ordered orbitals, allows the correlation space to be split into groups of CSFs, where each group is obtained 
by spin-angular coupling preserving orbital de-excitations from a generating CSFs with the orbitals in the symmetry ordered set \ytl{having the principal quantum numbers}  at their highest \mrg{value}.
The generating CSFs are of 4 types. Type~1: generating CSF has one orbital in the symmetry ordered set with the principal quantum number at its highest. Type~2: generating CSF has two orbitals of different symmetries 
in the symmetry ordered set, with
the principal quantum numbers at their highest. Type~3: generating CSF has two orbitals of the same symmetry in the symmetry ordered set, with
the principal quantum numbers at their highest and second highest. Type~4: generating CSF has a doubly occupied orbital 
in the symmetry ordered set, with the principal quantum number at its highest. Taken together, a generating CSF along with the orbital de-excitation rule is referred to as a 
configuration state function generator (CSFG), which generates a set of CSFs. In the remaining sections, the notions of CSFs obtained from a generating CSF by 
applying the de-excitation rules and CSFs generated by a CSFG will be used interchangeably. With the same meaning, we will also refer to  CSFs obtained by expanding the CSFGs or to  CSFs spanned by the CSFGs.

Let the labeling space, as part of a concrete example for a four electron system illustrating the concepts, be generated by SDTQ excitations from 
the $1s^22s2p$ reference configuration to the labeling ordered orbital set in eq. (\ref{eq:label})
%\[
%\Big\{1s,2s,2p\mbox{-},2p,3s,3p\mbox{-},3p,3d\mbox{-},3d\Big\}
%\]
and the
correlation space by SD excitations from $1s^22s2p$ to the symmetry ordered orbital set in eq. (\ref{eq:sym})
%\[
%\Big\{4s,5s,6s,7s,4p\mbox{-},5p\mbox{-},6p\mbox{-},7p\mbox{-},4p,5p,6p,7p,4d\mbox{-},5d\mbox{-},6d\mbox{-},7d\mbox{-},
%4d,5d,6d,7d\Big\} .
%\]
The groups of CSFs in the correlation space are obtained by the generating CSFs along with the de-excitation rules.
An example of a generating CSF of type 1 is given by
\begin{Verbatim}[fontsize=\footnotesize]
  1s ( 2)  2s ( 1)  7p-( 1)
               1/2      1/2
                           0-
\end{Verbatim}
The $7p\mbox{-}$ orbital is in the symmetry ordered set with the
principal quantum number at its highest. A group of four CSFs follows by spin-angular coupling preserving de-excitations of $7p\mbox{-}$ within the  $p\mbox{-}$ symmetry ordered orbital set
\begin{Verbatim}[fontsize=\footnotesize]
  1s ( 2)  2s ( 1)  4p-( 1)
               1/2      1/2
                           0-
              ...
	
  1s ( 2)  2s ( 1)  7p-( 1)
               1/2      1/2
                           0-
\end{Verbatim}
An example of a generating CSF of type 2 is given by
\begin{Verbatim}[fontsize=\footnotesize]
  1s ( 2)  7s ( 1)  7p-( 1)
               1/2      1/2
                           0-
\end{Verbatim}
The $7s$ and $7p\mbox{-}$ orbitals are in the symmetry ordered set with the principal quantum
numbers at their highest. A group of 16 CSFs follows by spin-angular coupling preserving de-excitations of $7s$ and $7p\mbox{-}$  within, respectively, the
$s$ and $p\mbox{-}$ symmetry ordered orbital sets
\begin{Verbatim}[fontsize=\footnotesize]
  1s ( 2)  4s ( 1)  4p-( 1)
               1/2      1/2
                           0-
  1s ( 2)  4s ( 1)  5p-( 1)
               1/2      1/2
                           0-
	            ...
	
  1s ( 2)  7s ( 1)  7p-( 1)
               1/2      1/2
                           0-
\end{Verbatim}
where the principal quantum number of the rightmost orbital moves more rapidly than the one for the second rightmost orbital. The generation of groups of CSFs from generating CSFs of types 3 and 4 works in the same way, and concrete examples are given in \cite{CSFG}. 
In the above example for $1s^22s2p$ there are 
altogether 200 generating CSFs (of which we have \mbox{shown two} \mrg{above}) for the $J=0$ odd parity block, generating in total 1200 CSFs that make up the correlation space.
The number of generating CSFs is independent of the size of the set of symmetry ordered orbitals. 
Increasing the set of symmetry ordered orbitals with three orbitals of each symmetry 
up to 
$10s,10p\mbox{-},10p,10d\mbox{-},10d$
gives the same 200 generating CSFs, but this time spanning
in total 7844 CSFs. At this point, it is important to \mrg{emphasize}
that the labeling CSFs together with the CSFs obtained by spin-angular coupling preserving 
de-excitations
from the generating CSFs span exactly the same functional space as the CSFs generated \mrg{through} the traditional way in {\sc Grasp2018}.
\subsection{The CSF file format in {\sc Graspg}}
The {\sc Grasp}2018 package is based on full lists of CSFs given in a file. {\sc Graspg} is based on  full lists of  CSFs in the labeling space, along with lists of generating CSFs.
Allowing SDTQ excitations from the $1s^22s2p$ reference configuration to the
labeling ordered orbital set in eq. (\ref{eq:label})
%\[
%\Big\{1s,2s,2p\mbox{-},2p,3s,3p\mbox{-},3p,3d\mbox{-},3d\Big\}
%\]
along with
SD excitations to the orbitals in the symmetry ordered set 
in eq. (\ref{eq:sym})
%\[
%\Big\{4s,5s,6s,7s,4p\mbox{-},5p\mbox{-},6p\mbox{-},7p\mbox{-},4p,5p,6p,7p,4d\mbox{-},5d\mbox{-},6d\mbox{-},7d\mbox{-},
%4d,5d,6d,7d\Big\} .
%\]
we obtain the {\sc Graspg} file displayed in table \ref{tab:CSFG1}. \clearpage

\begin{table}[h]
\caption{File with labeling and generating CSFs for the $J=0$ odd parity block obtained by the {\tt rcsfggenerate\_csfg} program. The orbitals in the correlation space are symmetry ordered.}\vspace{-3 mm}
\label{tab:CSFG1}
\hhrule
\footnotesize{

\begin{verbatim}
Core subshells:

Peel subshells:
  1s   2s   2p-  2p   3s   3p-  3p   3d-  3d   4s   5s   6s   7s
  4p-  5p-  6p-  7p-  4p   5p   6p   7p   4d-  5d-  6d-  7d-  4d
  5d   6d   7d
CSF(s):
  1s ( 2)  2s ( 1)  2p-( 1)     <-- start CSFs in labeling space
               1/2      1/2
                           0-

            ....


  3p-( 1)  3d-( 2)  3d ( 1)     <-- end CSFs in labeling space
      1/2        2      5/2
                  5/2      0-
  1s ( 2)  2s ( 1)  7p-( 1)     <-- start generating CSFs in 
               1/2      1/2         correlation space
                           0-

            ....


  2s ( 1)  2p-( 1)  7d-( 2)     <-- end generating CSF in 
      1/2      1/2        0         correlation space
                    0      0-
\end{verbatim}}
\hhrule
\end{table}

\noindent
The orbitals that are used to build the CSFs are listed under {\tt Peel subshells} heading. First \mrg{come} the orbitals in the labeling ordered set and then \mrg{come} the orbitals in the symmetry ordered set. The orbitals are followed by the
CSFs in the labeling space. These, in turn, are followed by the generating CSFs.
The CSFs are specified according to the conventions in 
{\sc Grasp2018}. 
 In addition to the {\sc Graspg} file, there is a file keeping track of the orbitals in the labeling ordered set. The latter file is displayed in table \ref{tab:nonsymorb}.  The de-excitation rules can be inferred from the orbitals given in the above two files.\clearpage

 \begin{table}[h]
\caption{File keeping track of the orbitals that are in the labeling ordered set, as obtained
by the {\tt rcsfggenerate\_csfg} program. \per{Here {\tt nonsym} is the number of orbitals that are not symmetry ordered, i.e., it is the number of labeing orbitals as listed on the line below.}  }
\vspace{-3 mm}
\label{tab:nonsymorb}
\hhrule
\footnotesize{

\begin{verbatim}
           9   = nonsym
  1s   2s   2p-  2p   3s   3p-  3p   3d-  3d
\end{verbatim}}
\hhrule
\end{table}

Specifying the generating CSFs, and not the full groups of CSFs, significantly reduces the amount of data to be stored on both disk and in memory. The size of the list is independent of the highest principal quantum numbers of the 
orbitals in the symmetry ordered set. It is thus especially 
attractive for calculations based on large orbital sets, with many correlation orbitals of the same symmetry.
\subsection{Spin-angular integrations and matrix elements based on CSFGs}\label{sec:saCSFG}
By using generating CSFs and de-excitation rules, 
matrix elements can be computed based on a limited number of spin-angular integrations. As an illustration, we compute the matrix elements of the Dirac-Coulomb Hamiltonian for a 
list in {\sc Graspg} format made up from one CSFs in the labeling space and two generating CSFs of types 1 and 2, respectively, as given in table \ref{tab:CSFG3}. \clearpage

\begin{table}[h]
\caption{List of CSFs in  {\sc Graspg} format. There is one CSF in the labeling space and two generating CSFs of types 1 and 2, respectively.\vspace{-3 mm}}
\label{tab:CSFG3}
\hhrule
\footnotesize{
\begin{verbatim}
Core subshells:

Peel subshells:
  1s   2s   2p-  2p   3s   3p-  3p   3d-  3d   4s   5s   6s   7s
  4p-  5p-  6p-  7p-  4p   5p   6p   7p   4d-  5d-  6d-  7d-  4d
  5d   6d   7d
CSF(s):
  1s ( 2)  2s ( 1)  2p-( 1)    <-- CSF in labeling space
               1/2      1/2
                           0-
  1s ( 2)  2s ( 1)  7p-( 1)    <-- generating CSF type 1
               1/2      1/2
                           0-													
  1s ( 2)  7s ( 1)  7p-( 1)    <-- generating CSF type 2
               1/2      1/2
                           0-
\end{verbatim}
}
\hhrule
\end{table}
\subsubsection{CSF in labeling space and CSFs obtained by generating CSF of \mbox{type 1}}
To compute the Dirac-Coulomb matrix elements between the CSF in the labeling space and {\em all} the CSFs obtained by the first generating CSF and the de-excitation rule, 
there are  calls to the spin-angular routines of the  {\tt librang90} library  of {\sc Grasp2018}  \cite{GRASPGG} to resolve the matrix element involving the generating CSF into spin-angular coefficients and radial integrals 							
\begin{Verbatim}[fontsize=\footnotesize]
    1.000000000 I(2p-,7p-)
    1.000000000 R0(2s 2p-,2s 7p-)
   -0.333333333 R1(2s 2s ,7p-2p-)
    2.000000000 R0(1s 2p-,1s 7p-)
   -0.333333333 R1(1s 1s ,7p-2p-)
\end{Verbatim}
Values of the radial integrals are read from memory. After multiplications with the angular coefficients, all the contributions are summed to form the matrix element. Each of the matrix elements involving the remaining 3 CSFs generated by the CSFG follow by keeping the spin-angular coefficients
from above, multiplying with radial integrals involving orbitals obtained by the appropriate de-excitation of $7p\mbox{-}$ within the
$p\mbox{-}$ symmetry ordered set and summing the contributions.
\subsubsection{CSFs generated by generating CSF of type 1}  
Spin-angular integration resolves the diagonal Dirac-Coulomb matrix element of the generating CSF into
\begin{Verbatim}[fontsize=\footnotesize]
    2.000000000 I(1s ,1s )
    1.000000000 I(2s ,2s )
    1.000000000 I(7p-,7p-)
    1.000000000 R0(2s 7p-,2s 7p-)
   -0.333333333 R1(2s 2s ,7p-7p-)
    1.000000000 R0(1s 1s ,1s 1s )
    2.000000000 R0(1s 2s ,1s 2s )
   -1.000000000 R0(1s 1s ,2s 2s )
    2.000000000 R0(1s 7p-,1s 7p-)
   -0.333333333 R1(1s 1s ,7p-7p-)
\end{Verbatim}
Just as above, the computer fetches the radial integrals from the memory, performs the multiplications and sums the contributions.
Each of the remaining diagonal matrix elements of the CSFs generated by the CSFG follow
by keeping the spin-angular coefficients, multiplying with  radial integrals involving orbitals obtained by appropriate de-excitations
of  $7p\mbox{-}$. All the off-diagonal matrix elements between CSFs generated by the CSFG follow
by keeping the spin-angular coefficients of the terms involving $7p\mbox{-}$ orbitals, multiplying with radial integrals involving orbitals obtained by   de-excitations of the leftmost and rightmost $7p\mbox{-}$ orbitals and summing the contributions. The appropriate de-excitations depend 
on the principal quantum number of the symmetry ordered $p\mbox{-}$ orbital in the left- and right-hand interacting CSFs. 
\subsubsection{CSFs generated by generating CSFs of type 1 and 2} 
Spin-angular integration resolves the Dirac-Coulomb matrix element between the two generating CSFs into
\begin{Verbatim}[fontsize=\footnotesize]
    1.000000000 I(2s ,7s )
    1.000000000 R0(2s 7p-,7s 7p-)
   -0.333333333 R1(2s 7s ,7p-7p-)
    2.000000000 R0(1s 2s ,1s 7s )
   -1.000000000 R0(1s 1s ,2s 7s )
\end{Verbatim}
The computer fetches the radial integrals from the memory, performs the multiplication and sums the result. The remaining matrix elements between CSFs for which the symmetry ordered $p\mbox{-}$ orbital for the left hand CSF is the same as for the right hand CSF  follow directly by 
keeping the spin-angular coefficients, multiplying with radial integrals involving orbitals obtained by the appropriate de-excitations of $7s$ and $7p\mbox{-}$.
In cases where the symmetry ordered $p\mbox{-}$ orbital for the left-hand CSF has a different principal quantum number than the 
one for the right-hand CSF, the matrix elements follow by keeping 
the spin-angular coefficients of the terms involving both the $7s$ and $7p\mbox{-}$ orbitals, multiplying with radial integrals involving orbitals obtained by  appropriate de-excitations of the $7s$ orbital and leftmost and rightmost $7p\mbox{-}$ orbitals and summing the contributions.

The above simple examples illustrate the method. For more examples, see \cite{CSFG}. Whatever combination of generating CSFs, one
never needs to perform more than three spin-angular integrations,
from which all the matrix elements between the CSFs obtained by the generating CSFs and the de-excitation rule follow by keeping the spin-angular coefficients
and multiplying with radial integrals involving orbitals obtained by appropriate de-excitations. 
The efficiency of the method relies on the fact that it keeps the spin-angular integrations needed to compute the full matrix at an absolute minimum and trades them with the 
fast process of orbital de-excitation.

\subsubsection{Spin-angular integration for the Breit interaction}
The use of CSFGs reduces the computational load for
determining the matrix elements of the Dirac-Coulomb Hamiltonian.
The same idea can be applied to the Breit interaction. The latter is more complicated, with the additional need to break down the interaction strength into a combination of different radial two-electron integrals, slowing down the computation. 
%The contribution of the Breit interaction to the total energy of an atomic system is growing with the increasing nuclear charge $Z$. Looking at the energy contribution per subshell, it is found that these decrease rapidly with the increasing orbital quantum number $l$ as well as with the increasing principal quantum number $n$, see for 
%example \cite{Breitorbital}. 
In order to speed things up, and take full advantage of the CSFGs, we may put restrictions on the Breit
interaction such that it is included only between CSFs built from a user defined subset of the full orbital set. This will reduce the number of Breit integrals and make sure that time is not spent on computing contributions that can be neglected \cite{CSFG}.
%The appropriate limitations of the orbital set depend on the atomic system and the desired accuracy, and should be balanced against the reduction in execution time. 

\subsection{Computing and storing spin-angular coefficients in {\sc Graspg} \label{sec:saCSFGcalc}}
In {\sc Graspg}, the Dirac-Coulomb matrix elements are resolved into spin-angular coefficients and pointers to radial one- and two-electron integrals using the {\tt rangular\_csfg\_mpi} program. The spin-angular coefficients together with pointers to the corresponding one- and two-electron integrals (with all possible tensor ranks \emph{k}) are written to the unformatted files {\tt csfg\_mcp.31} and {\tt csfg\_mcp.32}, see \cite{GRASP2018_Man} section 2. The file {\tt csfg\_mcp.30} records the interacting CSF pairs with their indices in the list of CSFs in {\sc Graspg} format, whereas the corresponding file {\tt mcp.30} in {\sc Grasp2018} stores the sparse structure of Hamiltonian matrix. The latter is now instead generated internally during the MCDHF calculation using the  {\tt rmcdhf\_csfg\_mpi} program. Due to the fact that the spin-angular integrations are performed only between the generating CSFs, the execution time and the amount of data for {\tt rangular\_csfg\_mpi} are significantly reduced  compared to {\sc Grasp2018}, even by factors of exceeding 100, as seen in sections \ref{sec:FEXV}, \ref{sec:scaling}, and the examples given in \ref{sec:scripts}.

%\subsection{Restrictions on the Breit interaction}
%Above, we have shown how the use of CSFGs reduces the computational load for
%determining the matrix elements of the Dirac-Coulomb Hamiltonian.
%The same idea can be applied to the Breit interaction. The latter is more complicated, with the additional need to break down the interaction strength into a combination of different radial two-electron integrals, slowing down the calculation. 
%The contribution of the Breit interaction to the total energy of an atomic system is growing with the increasing nuclear charge $Z$. Looking at the energy contribution per subshell, it is found that these decrease rapidly with the increasing orbital quantum number $l$ as well as with the increasing principal quantum number $n$, see for 
%example \cite{Breitorbital}. In order to speed up the calculations, and take full advantage of the CSFGs, we may put restrictions on the Breit
%interaction such that it is included only between CSFs built from a user defined subset of the full orbital set. This will reduce the number of Breit integrals and make sure that time is not spent to compute contributions that can be neglected. 
%The appropriate limitations of the orbital set depend on the atomic system and the desired accuracy, and should be balanced against the reduction in execution time. 
\subsection{MCDHF calculations in {\sc Graspg}}\label{sec:mcdhfCSFG}
The MCDHF calculations in {\sc Graspg} are performed in the same way as in {\sc Grasp2018}. The matrix elements are resolved into spin-angular coefficients and pointers to radial one- and two-electron integrals using the
{\tt rangular\_csfg\_mpi} program.
%\cyc{\sout{The spin-angular coefficients together with pointers to the corresponding one- and two-electron integrals are written to the unformatted {\tt csfg\_mcp.XXX} 
%files, see \cite{GRASP2018_Man} section 2. Due to the fact that we spin-angular integration is performed only between the generating CSFs, the amount of data is much smaller than for {\sc Grasp2018}. moved to section \ref{sec:saCSFGcalc}}}
At the next stage, the 
{\tt rmcdhf\_csfg\_mpi} program loads the list of CSFs in {\sc Graspg} format, the initial estimates of the radial orbitals, and the contents of the 
{\tt csfg\_mcp.XXX} files, rather than the full {\tt mcp.XXX} files of section \ref{sec:mcp}, into arrays, with a substantial reduction of the memory requirements compared to  {\sc Grasp2018}. 
Matrix elements between CSFs, as well as \mrg{direct} and exchange potentials of the coupled integro-differential equations for the 
radial orbitals, are inferred from 
spin-angular data between generating CSFs and applying the de-excitation rules according to the prescription in section \ref{sec:saCSFG}. Unlike the old {\tt rmcdhf} program, the {\tt rmcdhf\_csfg\_mpi} program evaluates all the needed radial integrals and saves them in memory before computing the Hamiltonian matrix.
The mixing coefficients of the CSFs, obtained by diagonalizing the Hamiltonian matrix, are with respect to the full list of CSFs generated by the CSFGs. 

Besides the significant memory reduction thanks to using the CSF list in {\sc Graspg} format, the execution time of {\tt rmcdhf\_csfg\_mpi} is also substantially reduced due to inclusion of the improvements presented in \cite{GRASPCPU}. \per{Specifically}, the algorithm to accumulate all the needed \mrg{direct} and exchange potentials of the coupled integro-differential equations is redesigned, and is more effective than that of the corresponding {\sc Grasp2018} program. The potentials are simultaneously obtained before the computations of the Lagrange multipliers. Some subroutines for diagonalizing the Hamiltonian matrix are additionally parallelized, as well as those for computing the potentials. As shown in \cite{GRASPCPU},  and also in section \ref{sec:FEXV} and \ref{sec:scripts}, these improvements could speed up the MCDHF calculations by an order of magnitude or more, compared to the {\sc Grasp2018} ones.  

\subsection{CI calculations in {\sc Graspg}}\label{sec:rciCSFG}
The CI calculations in {\sc Graspg} are performed in the same way as in {\sc Grasp2018}, but with the list 
of CSFs in {\sc Graspg} format with generating CSFs. 
The {\tt rci\_csfg\_mpi} 
program starts by evaluating the needed radial integrals and effective interaction
strengths and saves them in memory.
Matrix elements between CSFs are inferred from the
spin-angular data between generating CSFs and applying the de-excitation rules according to the prescription in section \ref{sec:saCSFG}. In the {\sc Graspg}
implementation, we may
put restrictions on the Breit interaction such that it is included only
between CSFs built from a user defined subset of the full orbital set.
Just as for {\sc Grasp2018}, the matrix elements are saved in the {\tt rci.res} files.
The computation of the Hamiltonian matrix is followed  by a call to the Davidson eigenvalue solver \cite{Davidson}
 to give the mixing coefficients of the CSFs for the targeted states. 
The mixing coefficients are with respect to the full list of CSFs generated by the CSFGs.
\subsection{Condensation}\label{sec:acc}
SD excitations from an MR to an orbital set often lead to a large number of CSFs, many of which 
have small mixing coefficients and negligible contributions to the total energy or to computed expectation values or transition amplitudes.
Removing CSFs with small mixing coefficients is known as ``condensation''~\cite{COND1}. 
Ideally, one wants 
to predict the CSFs that are unimportant and can be discarded for large MCDHF or CI calculations based 
on the information from smaller  calculations, see for example \cite{Deep}.
As shown in~\cite{CSFG,iCI}, the ordered structure of the correlation space resulting from
the use of CSFGs supports such a prediction. In table~\ref{tab:condense} we have
the full list of CSFs to the left and the corresponding CSFGs to the right. 
\mrg{In this table, there are in total $M$ CSFGs. 
Of these, CSFG$^i(n_i^{max},m_i^{max})$ is the $i$th one. 
CSF$^i_1, \ldots,$CSF$^i_{N_i}$ are the $N_i$ CSFs spanned by this particular CSFG. Each CSFG has one or two orbitals in the symmetry ordered orbital set, with the principal quantum number at their highest (e.g. $7s$ and $7p\mbox{-}$ on page 13). These numbers are indicated by the two integers $n_i^{max}$ and $m_i^{max}$ in parentheses. 
}
%
% CSFG^1(n_1^{max},m_1^{max}) is the first CSFG and
% CSF^1_1,...,CSF^1_{N_1} are the CSFs spanned by this CSFG.
%
% CSFG^M(n_M^{max},m_ M^{max}) is the last CSFG and
% CSF^M_1,...,CSF^M_{N_ M} are the CSFs spanned by this CSFG.
%
% There are in total M CSFGs.
%
% Everything refers to the situation before condensation.
%
% Each CSFG has one or two orbitals in the symmetry ordered orbital set with the principal quantum number at their highest, e,g, 7s, and 7p- on page 13.  In CSFG^1(n_1^{max},m_1^{max}) the notation n_1^{max} and m_1^{max}) are the (maximal) principal quantum numbers of the two symmetry ordered orbitals. 
%
% The two numbers need not be the same, if the symmetry ordered orbitals have s and p- symmetry we may for example have principal quantum number n_1^{max} = 8 for the s symmetry and m_ 1^{max} = 7 for p-.  
%
% I agree that the notation may not be the best and that one maybe need to include the symmetry of the symmetry ordered orbital in the notation.  Also, we have cases where we have only one orbital in the symmetry ordered set and this is not clear from the notation either.
%
The CSFGs
to the right include the information on the maximum principal quantum numbers of the orbitals in the symmetry ordered set.\clearpage

\begin{table}[h]
\caption{CSFs and corresponding CSFGs. The list of CSFs can be condensed by sorting the CSFGs according to their accumulated squared weights (mixing coefficients) and
retaining only the one's contributing to a predefined fraction of the total squared weight. The condensed list can be extended to larger orbital sets.\vspace{-3 mm}}
\label{tab:condense}
\hhrule
    
{\scriptsize
\[
\begin{array}{lll}
\mbox{CSF}_1 & \hspace{-0.2mm}\bigr \rceil    &   \\
%\mbox{CSF}_2 & \big|               &  \\
~~~:    & \Big|               &  \mbox{labeling space}  \\
\mbox{CSF}_N & \hspace{-0.2mm}\bigr \rfloor  & \medskip\\
\mbox{CSF}^1_1 &\hspace{-0.2mm}\bigr \rceil    &   \\
%\mbox{CSF}^1_2 & \bigg|               &  \\
~~~:    & \Big|               &  \mbox{CSFG}^1(n_1^{\mbox{\tiny max}},m_1^{\mbox{\tiny max}}) \\
\mbox{CSF}^1_{N_1} & \hspace{-0.2mm}\bigr \rfloor  & \medskip\\
%\mbox{CSF}^2_1 &\hspace{-0.2mm}\bigr \rceil    &  \\
%\mbox{CSF}^2_2 & \bigg|               &  \mbox{CSFG}^2(n_2^{\mbox{\tiny max}},m_2^{\mbox{\tiny max}}) \\
%~~~:    & \Big|               &  \\
%\mbox{CSF}^2_{N_2} & \hspace{-0.2mm}\bigr \rfloor  & \\ 
~~~    \vdots & & \medskip\\      
\mbox{CSF}^M_1 &\hspace{-0.2mm}\bigr \rceil    &  \\
%\mbox{CSF}^M_2 & \bigg|               &  \mbox{CSFG}^M(n_M^{\mbox{\tiny max}},m_M^{\mbox{\tiny max}}) \\
~~~:    & \Big|               &  \mbox{CSFG}^M(n_M^{\mbox{\tiny max}},m_M^{\mbox{\tiny max}}) \\
\mbox{CSF}^M_{N_{\!M}} & \hspace{-0.2mm}\bigr \rfloor  & \\ 
\end{array}
\]
}
\hhrule
\end{table}
\noindent
Perform an MCDHF or a CI calculation targeting one state -- the treatment is extended to the
multistate case by taking relevant means.
Assign each CSFG a squared
weight equal to the sum of the squared weights (mixing coefficients) of the CSFs generated by the CSFG. Include all squared 
weights from the CSFs in the labeling space, sort the
squared weights of the \mbox{CSFGs}, and accumulate to a predefined fraction, e.g., 0.99999999, of 1. This procedure
gives a number of surviving CSFGs along with unimportant \mbox{CSFGs} that can be discarded. Now, the surviving CSFGs
are used for larger MCDHF or CI calculations based on extended sets of symmetry ordered orbitals. 
For an extended set with, say, $k$ added orbital layers, this amounts to a CSF list in {\sc Graspg} format, 
for which the following replacements
\begin{equation}\label{stepup}
\mbox{CSFG}(n^{\mbox{\scriptsize max}},m^{\mbox{\scriptsize max}}) \rightarrow
\mbox{CSFG}(n^{\mbox{\scriptsize max}}+k,m^{\mbox{\scriptsize max}}+k)
\end{equation}
have been done for the surviving CSFGs.  
\section{Installation of {\sc Graspg} and merging with {\sc Grasp}2018}

%\subsection{Downloading and installing {\sc Graspg}}
\noindent

The {\sc Grasp}2018 package~\cite{GRASP2018} consists of libraries, application programs, and tools written in Fortran 95 and adapted
to run in parallel under MPI, a language-independent communication protocol. The package comes with a detailed manual describing code organization, compilation, and file flow \cite{GRASP2018_Man}. The manual also includes test runs, examples, and script files.
 {\sc Graspg}2018 can be downloaded from
GitHub: 
%\href{https://github.com/compas}
\url{https://github.com/compas}. The main directory, \per{ {\tt grasp-master}}, of the downloaded {\sc Grasp}2018 package contains the following
directories:

\begin{tabular}{ll}
          & \\
{\tt bin} & directory where, after compilation, the executables reside \\
{\tt lib} & directory where, after compilation, the static library \\
          & archives reside  \\
{\tt src} & directory with the subdirectories {\tt appl}, containing \\
          & the source code for the application programs, {\tt lib}, \\
          & containing the source code for the libraries and {\tt tool}, \\
          & containing the source code for the tools.\\
{\tt grasptest} & directory containing scripts for all the test runs and \\
          & examples in the corresponding manual \cite{GRASP2018_Man}.\\
          & \\
\end{tabular}
{\sc Graspg} has been developed as an extension of {\sc Grasp}2018 and can be downloaded from
GitHub: 
%\href{https://github.com/compas}
\url{https://github.com/compas}. 
{\sc Graspg} contains the following
directories:

\begin{tabular}{ll}
          & \\
%{\tt bin} & directory where, after compilation, the executables reside \\
%{\tt lib} & directory where, after compilation, the static library \\
%          & archives reside  \\
{\tt srcg} & directory with the subdirectories {\tt appl}, containing \\
          & the source code for the application programs, {\tt lib}, \\
          & containing the source code for the libraries and {\tt tool}, \\
          & containing the source code for the tools.\\
{\tt graspgtest} & directory containing scripts for all the test runs and \\
          & examples to be described in later section of this article.\\
          & \\
\end{tabular}

The steps below, under the $\texttt{bash}$ shell, should be followed to ensure a proper installation of \cyc{{\sc Graspg}} and merger into {\sc Grasp}2018. 
The installation procedure  assumes that the {\sc Grasp}2018 package has already been installed. \\

\begin{enumerate}
\item Go to the main directory \per{ {\tt grasp-master}} of the installed {\sc Grasp}2018 package. Type
  $$\texttt{source make-environment\_xxx}$$ where {\tt xxx} is the compiler name (see the {\tt README} file of the {\sc Grasp}2018 package). The {\sc Grasp}2018 environment variables are now set.
\item Copy the {\tt graspg} package, downloaded from GitHub, to the \per{ {\tt grasp-master}} directory. 
  Untar the package, the directories {\tt srcg} and {\tt graspgtest} will now appear.
\item In the \per{ {\tt grasp-master/srcg}} directory, execute the installation by issuing the commands
  $$\texttt{make clean}$$ $$\texttt{make}$$ This will 
generate static library
archives that reside in the  \per{ {\tt grasp-master/lib}} directory. It will also generate executable program files that 
reside  in the \per{ {\tt grasp-master/bin}} directory. {\sc Graspg} is now fully merged with the {\sc Grasp}2018 package.
\end{enumerate}
After installation of {\sc Graspg} the following five new libraries should be found in the \per{ {\tt grasp-master/lib}} directory:

\begin{tabular}{lll}
 & & \\
{\tt libmod\_csfg.a} & {\tt lib9290\_csfg.a} & {\tt librang90\_csfg.a} \\
 {\tt libme\_csfg.a} & {\tt libdvdmpi.a} & \\
 & & \\
\end{tabular}

\noindent
The following ten executable application programs should be found in the \per{ {\tt grasp-master/bin}} directory, where
the extension {\tt \_mpi} indicates that the executable can be run in parallel  under MPI:

\begin{tabular}{lll}
 & &  \\
{\tt jj2lsj\_csfg} & {\tt rangular\_csfg\_mpi} & {\tt rci\_block\_csfg\_mpi } \\
{\tt rci\_csfg\_mpi} & {\tt rcsfggenerate\_csfg} & {\tt rmcdhf\_csfg\_mpi} \\
 {\tt rmixaccumulate\_csfg} & {\tt rwfnestimate\_csfg} & {\cyc{\tt rcsfginteract\_csfg}}   \\
 {\cyc{\tt rcsfgzerofirst\_csfg}} & & \\
 & & \\
\end{tabular}

\noindent
In the same directory, the following seven executable tools should also be found:

\begin{tabular}{lll}
 & &   \\
{\tt rcsfgexpand\_csfg} & {\tt rcsfgextend\_csfg} & {\tt rcsfgsplit\_csfg} \\
{\tt rdistHmatrix\_csfg}  & {\tt rcsfgblocksplit\_csfg} & {\tt rsave\_csfg}  \\
{\cyc{\tt rlevels\_csfg}} & &  \\
 & & \\
\end{tabular}

\noindent
The subdirectory \per{ {\tt grasp-master/graspgtest}} lists a number of 
script files for illustration 
of the {\sc Graspg} program usage to obtain energy structure of atoms and ions, see \ref{sec:scripts}.

\section{Description of the {\sc Graspg} application programs and tools}
Below is a brief description of the application programs and tools, ordered in categories depending on what they do.\medskip\\
Routines that generate and manipulate lists of CSFs in {\sc Graspg} format, i.e., lists of CSFs in the labeling space along with generating CSFs, see also the {\sc Grasp2018} manual \cite{GRASP2018_Man} sections 4, 5 and 7.1:
\begin{enumerate}
%\item {\tt rcsfexcitation}    --  generate excitation input to {\tt rcsfgenerate}
\item {\tt rcsfggenerate\_csfg} $-$ generate a list of CSFs in {\sc Graspg} format based on rules for orbital substitutions.
\item {\tt rcsfgsplit\_csfg} $-$ split a list of CSFs in {\sc Graspg} 
format into a number of lists, with the generating CSFs built from different user defined sets of symmetry ordered orbitals.
\item {\tt rcsfgblocksplit\_csfg} $-$ split a list of CSFs in {\sc Graspg} 
format into a number of lists, one for each $Jp$-block.
\item {\tt rcsfgextend\_csfg} $-$  extend the set of symmetry ordered orbitals for a list of CSFs in {\sc Graspg} format.
%\item {\tt rcsfinteract\_omp}  $-$  reduce a list of CSFGs by retaining only CSFGs that interact with CSFGs of a reference list. \cyc{\sout{????}}
\item {\tt rcsfgexpand\_csfg} $-$ expand a list of CSFs in {\sc Graspg} format, to a list of CSFs in ordinary {\sc Grasp2018} format.
%\item {\tt rcsfzerofirst\_omp} $-$ rearrange a list of CSFGs in such a way that the most important CSFGs are listed at the beginning, defining the zero-order space, and the less important are listed at the end, defining the first-order space, see TP \S ???.
\end{enumerate}
Programs to perform MCDHF calculations based on CSF lists in {\sc Graspg} format, with the method for spin-angular integration described in \mbox{section \ref{sec:saCSFG}:}
\begin{enumerate}
\item  {\tt rangular\_csfg\_mpi} -- perform spin-angular integration and compute angular coefficients 
based on a CSF list in {\sc Graspg} format. Includes options to treat some interactions perturbatively.
\item  {\tt rmcdhf\_csfg\_mpi} -- determine radial parts of the orbitals along with mixing coefficients with respect to the full list of CSFs
generated by the CSFGs in a relativistic self-consistent-field (SCF) procedure.
\end{enumerate}
Programs and tools to perform CI calculations with transverse photon (Breit) interaction and vacuum polarization and self-energy (QED) corrections based on CSF lists in {\sc Graspg} format with the method for spin-angular integration described in \mbox{section \ref{sec:saCSFG}.}
\begin{enumerate}
\item  {\tt rci\_csfg\_mpi} -- perform a CI calculation based on a CSF list in {\sc Graspg} format.
Mixing coefficients are given with respect to the full list of CSFs generated by the CSFGs. Includes options to treat some interactions perturbatively.
\item  {\tt rci\_block\_csfg\_mpi} -- compute, using many MPI processes, the Hamiltonian matrix  for one $Jp$-block based on a CSF list in {\sc Graspg} format. 
In restart mode, after a redistribution of the {\tt rci.res} files using {\tt rdistHmatrix\_csfg}, the Hamiltonian is diagonalized using fewer MPI processes. 
Includes options to treat some interactions perturbatively.
\item  {\tt rdistHmatrix\_csfg} $-$  redistribute the {\tt rci.res} files holding the Hamiltonian matrix from an {\tt rci\_block\_csfg\_mpi} run using many MPI processes to a structure consistent with fewer MPI processes.
\end{enumerate}
{\tt jj2lsj\_csfg} $-$ convert a portion of the wave function expansion in $jj$-coupled
CSFs in {\sc Graspg} format to a basis of $LSJ$-coupled CSFs for labeling purposes, see \cite{GG4b,GG4,GG5,ATOMSJJLSJ}. Includes a feature to provide unique labels for all the considered states.\medskip\\
{\tt rmixaccumulate\_csfg} $-$ condense a CSF list in  {\sc Graspg} format based on accumulation to a specified fraction according to the prescription in section \ref{sec:acc}. Includes a feature to generate 
condensed CSF lists in  {\sc Graspg} format, built on extended sets of symmetry ordered orbitals, which amounts to \mbox{{\em a priori}} condensation. \medskip\\
\cyc{{\tt rlevels\_csfg} $-$ list the levels in a series of mixing files, in the order of increasing energy and report levels in cm$^{-1}$
 relative to the lowest. If the {\tt jj2lsj\_csfg} program has been run, the levels are given in $LSJ$-coupling notation.} \medskip\\
\cyc{{\tt rcsfginteract\_csfg} $-$ reduce a CSFs list in  {\sc Graspg} by retaining only CSFs that interact with CSFs of a reference list, see \cite{GRASP2018_Man} Section 5.5.} \medskip\\
\cyc{{\tt rcsfgzerofirst\_csfg} $-$ rearrange a CSFs list in {\sc Graspg} format in such a way that the most important CSFs are listed at the beginning, defining the zero-order space, and the less important are listed at the end, defining the first-order space, see \cite{GRASP2018_Man} Section 2.8.} 
%\end{enumerate}

\section{File Naming Convention}
In the same way as for  {\sc Grasp}2018,
the passing of information between different programs in {\sc Graspg} is done through files,  where the extension 
of the file names indicates the content of the file. {\sc Graspg} uses the same extensions as {\sc Grasp}2018.
In addition, it has four extensions related to lists of CSFs in  {\sc Graspg} format.

\begin{table}[h!]
\caption{Additional file extensions for {\sc Graspg}.}
\centering
\label{tab:extension}
{\small
\begin{tabular}{llll} \hline 
Extension & Type of file\\
\hline
{\tt g}  & list of CSFs in {\sc Graspg} format \\
{\tt l}  & list of labeling  orbitals \\
{\tt rem}  & generating CSFs removed from list during condensation \\ 
{\tt nmax}  & list of generating CSFs from condensation with extended\\ 
            &  set of symmetry ordered orbitals\\
    \hline 
\end{tabular}}
\end{table}

\section{Program  and Data Flow}
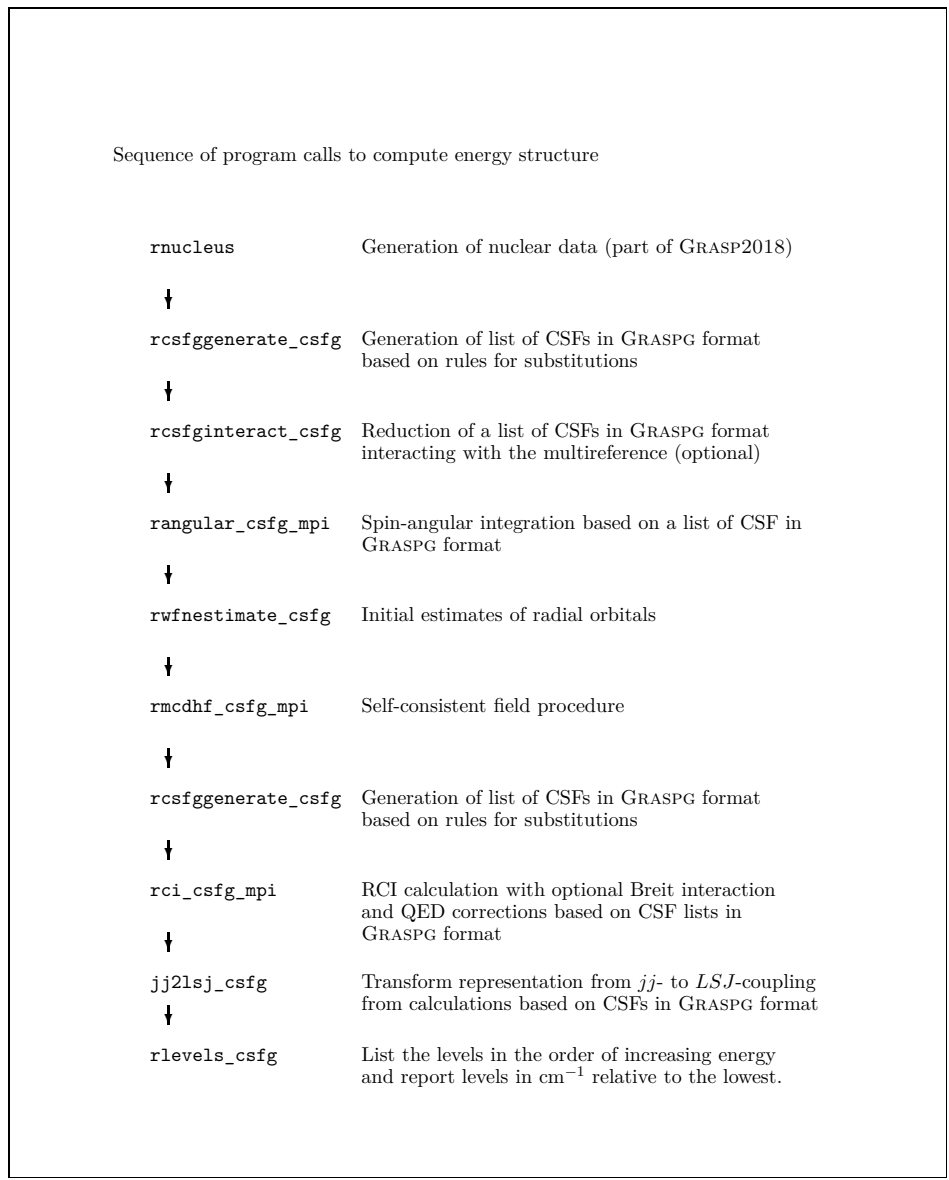
\begin{figure}
\resizebox{1.0\linewidth}{!}{
\fbox{
%\begin{picture}(450,630)(0,-100)
\begin{picture}(500,630)(0,-100)
%%\begin{picture}(500,550)(0,-60)
\thicklines
\put (50,450){Sequence of program calls to compute energy structure}
\put (70,400){{\tt rnucleus}}
\put (185,400){Generation of nuclear data (part of {\sc Grasp2018})}
\put (80,380){\vector(0,-1){10}}

%\put (100,450){{\tt rcsfexcitation}}
%\put (185,450){Specify multireference, orbital set and excitations}
%\put (110,430){\vector(0,-1){10}}

\put (70,350){{\tt rcsfggenerate\_csfg}}
\put (185,350){Generation of list of CSFs in {\sc Graspg} format} 
\put (185,338){based on rules for substitutions}

\put (80,330){\vector(0,-1){10}}
\put (70,300){\per{{\tt rcsfginteract\_csfg}}}
\put (185,300){Reduction of a list of CSFs in {\sc Graspg} format }
\put (185,288){interacting with the multireference (optional)}

\put (80,280){\vector(0,-1){10}}
\put (70,250){{\tt rangular\_csfg\_mpi}}
\put (185,250){Spin-angular integration based on a list of CSF in  }
\put (185,238){{\sc Graspg} format}
\put (80,230){\vector(0,-1){10}}

\put (70,200){{\tt rwfnestimate\_csfg}}
\put (185,200){Initial estimates of radial orbitals}
\put (80,180){\vector(0,-1){10}}

\put (70,150){{\tt rmcdhf\_csfg\_mpi}}
\put (185,150){Self-consistent field procedure}
\put (80,130){\vector(0,-1){10}}

\put (70,100){{\tt rcsfggenerate\_csfg}}
\put (185,100){Generation of list of CSFs in {\sc Graspg} format} 
\put (185,88){based on rules for substitutions}
\put (80,80){\vector(0,-1){10}}

\put (70,50){{\tt rci\_csfg\_mpi}}
\put (185,50){RCI calculation with optional Breit interaction}
\put (185,38){and QED corrections based on CSF lists in}
\put (185,26){{\sc Graspg} format}
\put (80,30){\vector(0,-1){10}}
\put (70,0){{\tt jj2lsj\_csfg}}
\put (185,0){Transform representation from $jj$- to $LSJ$-coupling}
\put (185,-12){from calculations based on CSFs in {\sc Graspg} format}

\put (80,-10){\vector(0,-1){10}}
\put (70,-40){\per{{\tt rlevels\_csfg}}}
\put (185,-40){List the levels in the order of
increasing energy }
\put (185,-52){and report levels in cm$^{-1}$ relative to the lowest. }

%\put (320,-30){Eval. of expect. values }
\end{picture}}
}
\caption{
Typical sequence of program calls to compute \per{and list} energy structure with the  {\sc Graspg} package.}
\label{fig:Typical_run}
\end{figure}

\begin{figure}
\resizebox{1.0\linewidth}{!}{
\fbox{
%\begin{picture}(450,630)(0,-100)
%\begin{picture}(500,630)(0,-100)
%\begin{picture}(500,350)(0,-100)
\begin{picture}(500,410)(0,-140)
\thicklines
\put (50,200){Sequence of program calls to compute energy structure} 
\put (60,150){{\tt rcsfggenerate\_csfg}}
\put (185,150){Generation of list of CSFs in {\sc Graspg} format} 
\put (185,138){based on rules for substitutions.}
\put (80,130){\vector(0,-1){10}}

\put (60,100){{\tt rci\_csfg\_mpi}}
\put (185,100){RCI calculation with optional Breit interaction}
\put (185,88){and QED corrections based on a CSF list in}
\put (185,76){{\sc Graspg} format}
\put (80,80){\vector(0,-1){10}}

\put (60,50){{\tt rmixaccumulate\_csfg}}
\put (185,50) {Condens the CSF list in {\sc Graspg} format based }
\put (185, 38){on accumulation to a specified fraction}
\put (80,30){\vector(0,-1){10}}

\put (60,0){{\tt rci\_csfg\_mpi}}
\put (185,0){RCI calculation with optional Breit interaction}
\put (185,-12){and QED corrections based on the condensed CSF}
\put (185,-24){list in {\sc Graspg} format}
\put (80,-20){\vector(0,-1){10}}

\put (60,-50){{\tt jj2lsj\_csfg}}
\put (185,-50){Transform representation from $jj$- to $LSJ$-coupling}
\put (185,-62){from the calculations in CSFG notation}
%\put (320,-30){Eval. of expect. values }

\put (80,-60){\vector(0,-1){10}}
\put (60,-90){\per{{\tt rlevels\_csfg}}}
\put (185,-90){List the levels in the order of
increasing energy }
\put (185,-102){and report levels in cm$^{-1}$ relative to the lowest. }

\end{picture}}
}
\caption{
Typical sequence of program calls to compute energy structure with the  {\sc Graspg} package based on a condensed list of CSFs.}
\label{fig:Typical_run3}
\end{figure}
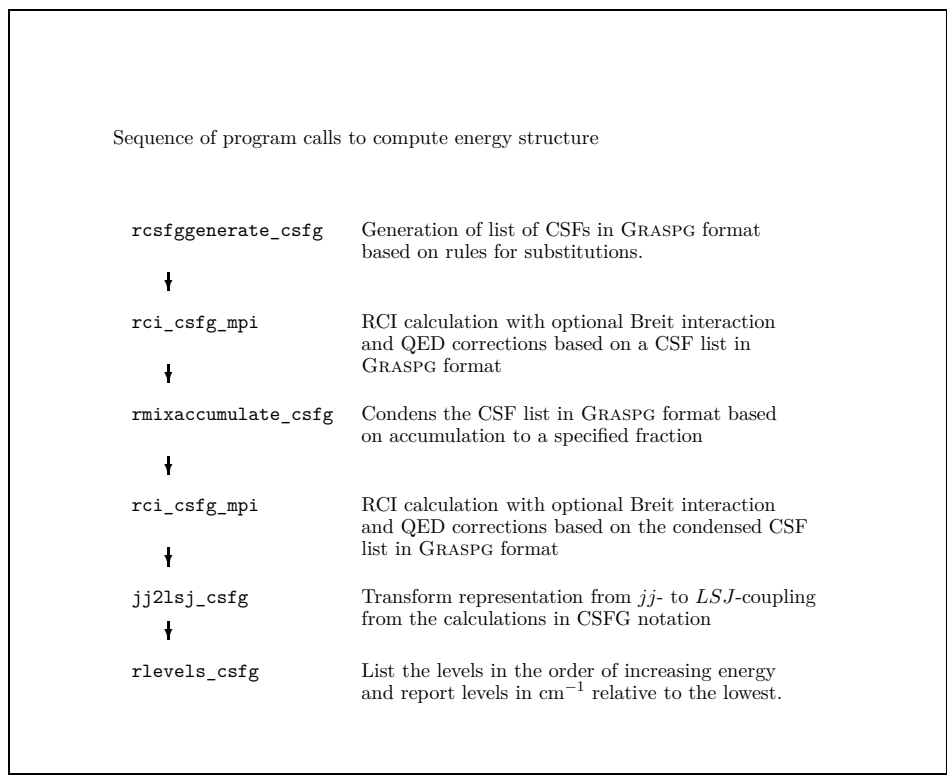

To perform a calculation, a number of programs need to be run in a predetermined order. 
Figures~\ref{fig:Typical_run}, \ref{fig:Typical_run2}, and \ref{fig:Typical_run3} show typical sequences of program calls to compute wave functions and energy structure with the programs of the {\sc Graspg} package.

\begin{figure}
\resizebox{1.0\linewidth}{!}{
\fbox{
%\begin{picture}(450,630)(0,-100)
%\begin{picture}(500,630)(0,-100)
%%\begin{picture}(500,350)(0,-100)
\begin{picture}(500,410)(0,-140)
\thicklines
\put (50,200){Sequence of program calls to compute energy structure} 
%CYC: \put (70,450){{\cyc{\tt rnucleus}}}
%CYC: \put (185,450){Generation of nuclear data}
%CYC: \put (80,430){\vector(0,-1){10}}

%\put (100,450){{\tt rcsfexcitation}}
%\put (185,450){Specify multireference, orbital set and excitations}
%\put (110,430){\vector(0,-1){10}}

%CYC: \put (70,400){{\tt rcsfggenerate}}
%CYC: \put (185,400){Generation of list of CSFs based on rules} 
%CYC: \put (185,388){for excitations}

%CYC: \put (80,380){\vector(0,-1){10}}
%CYC: \put (70,350){{\tt rcsfinteract}}
%CYC: \put (185,350){Reduction of a list to CSFs interacting with}
%CYC: \put (185,338){the multireference (optional)}

%CYC: \put (80,330){\vector(0,-1){10}}
%CYC: \put (70,300){{\tt rangular}}
%CYC: \put (185,300){Angular integration}
%CYC: \put (80,280){\vector(0,-1){10}}

%CYC: \put (70,250){{\tt rwfnestimate}}
%CYC: \put (185,250){Initial estimates of radial orbitals}
%CYC: \put (80,230){\vector(0,-1){10}}

%CYC: \put (70,200){{\tt rmcdhf}}
%CYC: \put (185,200){Self-consistent field procedure}
%CYC: \put (80,180){\vector(0,-1){10}}

\put (70,150){{\tt rcsfggenerate\_csfg}}
\put (185,150){Generation of list of CSFs in {\sc Graspg} format for a} 
\put (185,138){specific $Jp$-block based on rules for substitutions. }
\put (80,130){\vector(0,-1){10}}

\put (70,100){{\tt rci\_block\_csfg\_mpi}}
\put (185,100){Build the Hamiltonian matrix for the specified $Jp$-block}
\put (185,88){using a large number of MPI processes, e.g., the number of}
\put (185,76){available CPU cores. Matrix distributed in {\tt rci.res} files}
\put (80,80){\vector(0,-1){10}}

\put (70,50){{\tt rdistHmatrix\_csfg}}
\put (185,50){Redistribute the {\tt rci.res} files to a structure consistent}
\put (185,38){with fewer MPI processes, typically between 16-32.}
\put (80,30){\vector(0,-1){10}}

\put (70,0){{\tt rci\_block\_csfg\_mpi}}
\put (185,0){RCI calculation in restart mode to diagonalize the}
\put (185,-12){Hamiltonian matrix with the number of MPI processes }
\put (185,-24){as given in {\tt rdistHmatrix\_csfg}.}
\put (80,-20){\vector(0,-1){10}}

\put (70,-50){{\tt jj2lsj\_csfg}}
\put (185,-50){Transform representation from $jj$- to $LSJ$-coupling}
\put (185,-62){from calculations based on CSFs in {\sc Graspg} format.}

\put (80,-60){\vector(0,-1){10}}
\put (70,-90){\per{{\tt rlevels\_csfg}}}
\put (185,-90){List the levels in the order of
increasing energy }
\put (185,-102){and report levels in cm$^{-1}$ relative to the lowest. }

%\put (320,-30){Eval. of expect. values }
\end{picture}}
}
\caption{Typical sequence of program calls to compute energy structure with the  {\sc Graspg} package, with the 
restart option and redistribution of the {\tt rci.res} files to
change the number of MPI processes for the diagonalization.}
\label{fig:Typical_run2}
\end{figure}
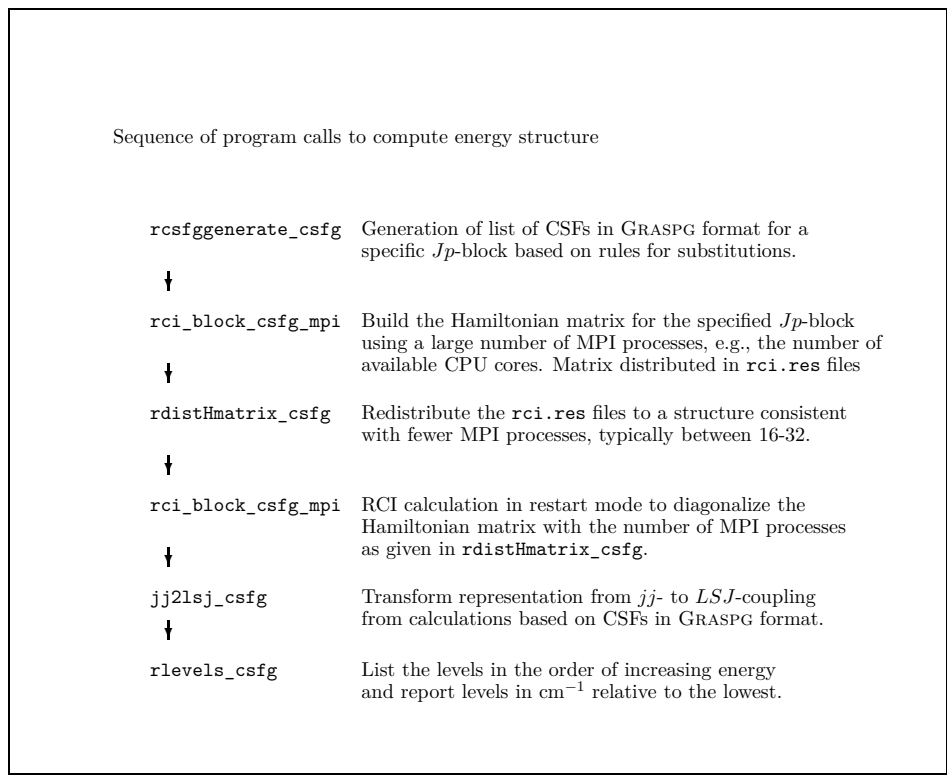

\clearpage
\section{Running the {\sc Graspg} application programs interactively}\label{sec:FEXV}
To demonstrate the use of the  {\sc Graspg} application programs and tools, we consider the energy structure for the
$3s^2,3p^2,3s3d,3d^2$ even and $3s3p,3p3d$ odd states in \mrg{magnesium-like} Fe XV. Taking  
the $3l3l'$ configurations as the MR, we determine the $n=7$ orbital layer
based on an MR-SD expansion accounting for valence-valence (VV) and core-valence (CV) electron correlation.
The $1s^2$ shell is closed in all CSFs.
The runs assume that the nuclear data file 
{\tt isodata} is available and that the user has performed layer-by-layer MCDHF calculations such that the orbitals up to $n = 6$ are available in the file {\tt n6.w}. In addition, the {\tt disks} file should be available defining the directories for the
temporary MPI files, see  {\sc Grasp2018} manual \cite{GRASP2018_Man} \mbox{section 6.4}.
\subsection{Generate list of CSFs in {\sc Graspg} format for the MCDHF calculation}
The application program {\tt rcsfggenerate\_csfg} is used to generate lists of CSFs in {\sc Graspg} format.
Note that one always need to start by generating CSFs in the labeling space. The orbitals used for the generation of the CSFs in the labeling space define
the labeling orbital set, the information on which is written to the {\tt rlabel.out} file. 
We start by generating CSFs in the even parity labeling space (we could equally well have started with the odd parity labeling space). As the next step 
we generate any additional CSFs in the even parity labeling space, as defined by the applied excitation rule 
and labeling orbital set, as well as generating CSFs in the correlation space, as defined by the applied excitation rule and extension of the orbital set beyond the labeling orbitals, i.e., including orbitals in the symmetry ordered set. As the final step we generate CSFs in the odd parity labeling space, as defined by the applied excitation rule 
and the labeling orbital set, as well as generating CSFs in the correlation space, as defined by the applied excitation rule and extension of the orbital set beyond the labeling orbitals.
%  { time nohup bash -c ./ek.sh ; } > output 2>&1 &

\begin{Verbatim}[fontsize=\footnotesize]
>>rcsfggenerate_csfg

 RCSFGGENERATE_CSFG
 This program generates a list of CSFG CSFs

 Configurations should be entered in spectroscopic notation
 with occupation numbers and indications if orbitals are
 closed (c), inactive (i), active (*) or has a minimal
 occupation e.g. 1s(2,1)2s(2,*)
 Outputfiles: rcsfg.out rlabel.out rcsfggenerate.log rcsfg.log clist.ref excitation  

 Select core
        0: No core
        1: He (       1s(2)                  =  2 electrons)
        2: Ne ([He] + 2s(2)2p(6)             = 10 electrons)
        3: Ar ([Ne] + 3s(2)3p(6)             = 18 electrons)
        4: Kr ([Ar] + 3d(10)4s(2)4p(6)       = 36 electrons)
        5: Xe ([Kr] + 4d(10)5s(2)5p(6)       = 54 electrons)
        6: Rn ([Xe] + 4f(14)5d(10)6s(2)6p(6) = 86 electrons)
>>1

 Generation of CSFs in labeling space
 Entered orbitals define the labeling orbital set
 Enter list of (maximum 1000) configurations. 
 End list with a blank line or an asterisk (*)

 Give configuration           1
>>2s(2,*)2p(6,*)3s(2,*)
 Give configuration           2
>>2s(2,*)2p(6,*)3p(2,*)
 Give configuration           3
>>2s(2,*)2p(6,*)3s(1,*)3d(1,*)
 Give configuration           4
>>2s(2,*)2p(6,*)3d(2,*)
 Give configuration           5
>>
 Give set of active orbitals, as defined by the highest principal quantum number
 in a comma delimited list in s,p,d etc order, e.g. 3s,3p,3d or 4s,4p,4d,4f or
 5s,5p,5d,5f,5g etc
>>3s,3p,3d
 Resulting 2*J-number? lower, higher (J=1 -> 2*J=2 etc.)
>>0,8
 Number of excitations (if negative number e.g. -2, correlation
 orbitals will always be doubly occupied)
>>0
 Generate more lists ? (y/n)
>>y
 Generation of CSFs in labeling space and, if entered orbitals extend 
 the labeling orbital set, also generating CSFs in correlation space
 Orbitals extending the labeling orbital set appear symmetry ordered
 Enter list of (maximum 1000) configurations. 
 End list with a blank line or an asterisk (*)

 Give configuration           1
>>2s(2,i)2p(6,5)3s(2,*)
 Give configuration           2
>>2s(2,i)2p(6,5)3p(2,*)
 Give configuration           3
>>2s(2,i)2p(6,5)3s(1,*)3d(1,*)
 Give configuration           4
>>2s(2,i)2p(6,5)3d(2,*)
 Give configuration           5
>>2s(2,1)2p(6,i)3s(2,*)
 Give configuration           6
>>2s(2,1)2p(6,i)3p(2,*)
 Give configuration           7
>>2s(2,1)2p(6,i)3s(1,*)3d(1,*)
 Give configuration           8
>>2s(2,1)2p(6,i)3d(2,*)
 Give configuration           9
>>
 Give set of active orbitals, as defined by the highest principal quantum number
 per l-symmetry, in a comma delimited list in s,p,d etc order, e.g.  5s,4p,3d
>>7s,7p,7d,7f,7g,7h,7i
 Resulting 2*J-number? lower, higher (J=1 -> 2*J=2 etc.)
>>0,8
 Number of excitations (if negative number e.g. -2, correlation
 orbitals will always be doubly occupied)
>>2
 Generate more lists ? (y/n)
>>y
 Generation of CSFs in labeling space and, if entered orbitals extend 
 the labeling orbital set, also generating CSFs in correlation space
 Orbitals extending the labeling orbital set appear symmetry ordered
 Enter list of (maximum 1000) configurations. 
 End list with a blank line or an asterisk (*)

 Give configuration           1
>>2s(2,i)2p(6,5)3s(1,*)3p(1,*)
 Give configuration           2
>>2s(2,i)2p(6,5)3p(1,*)3d(1,*)
 Give configuration           3
>>2s(2,1)2p(6,i)3s(1,*)3p(1,*)
 Give configuration           4
>>2s(2,1)2p(6,i)3p(1,*)3d(1,*)
 Give configuration           5
>>
 Give set of active orbitals, as defined by the highest principal quantum number
 per l-symmetry, in a comma delimited list in s,p,d etc order, e.g.  5s,4p,3d
>>7s,7p,7d,7f,7g,7h,7i
 Resulting 2*J-number? lower, higher (J=1 -> 2*J=2 etc.)
>>0,8
 Number of excitations (if negative number e.g. -2, correlation
 orbitals will always be doubly occupied)
>>2
 Generate more lists ? (y/n)
>>n
    ....
    
 RCSFGGenerate_CSFG: Execution complete.
\end{Verbatim}
\noindent
The program produces three files:  {\tt rcsfg.out}, which contains the CSF list in {\sc Graspg} format, {\tt rlabel.out}, which keeps track of the labeling orbitals, and {\tt rcsfggenerate.log}, which is a log-file that mirrors the input.
In practical work, it is often convenient to edit the log-file and use this as input for additional runs of {\tt rcsfggenerate\_csfg}.
In this case the orbitals in the labeling space are 
\[
\Big\{1s,2s,2p\mbox{-},2p,3s,3p\mbox{-},3p,3d\mbox{-},3d\Big\}.
\]
In general, once the CSF list in {\sc Graspg} format has been generated for the largest intended orbital set for the study at hand,
one can employ the {\tt rcsfgsplit\_csfg} program to split it into lists using smaller orbital sets, compare  {\sc Grasp2018} manual \cite{GRASP2018_Man} section 7.1, see also the scripts presented in \ref{sec:scripts}. 
For example, the above generated list for the $n=7$ calculation can be split and used as input for the $n=4,\, 5,\, 6$ calculations. One can also employ the {\tt rcsfgextend\_csfg} program to extend the $n=7$ list to serve as input for larger calculations, e.g., for $n=8,9,\ldots$. \per{Note that when using the 
{\tt rcsfgextend\_csfg} program, the extension should be restricted to symmetries for which there are at least two orbitals in the symmetry ordered set of the original list.}
\subsection{Expand list to see the number of CSFs}
The output file {\tt rcsfg.out} contains labeling CSFs and generating CSFs in {\sc Graspg} format. To expand the list to a 
full list of CSFs in  {\sc Grasp2018} format, we apply the de-excitation rules to the generating CSFs. This is 
done by the program {\tt rcsfgexpand\_csfg}. The input files are {\tt rcsfg.inp} and {\tt rlabel.inp}, and we start
by copying the necessary files
\begin{Verbatim}[fontsize=\footnotesize]
>>cp rcsfg.out rcsfg.inp
>>cp rlabel.out rlabel.inp
\end{Verbatim}
We now run the expansion program
\begin{Verbatim}[fontsize=\footnotesize]
>>rcsfgexpand_csfg

 RCSFGEXPAND_CSFG
 This program expands a list of CSFGs
 Input files: rcsfg.inp, rlabel.inp
 Output file: rcsf.out
    ....

 block         1  NCSF(L) =        53  NCSF(G) =      1200  NCSF =      8579
 block         2  NCSF(L) =        49  NCSF(G) =      1214  NCSF =      8542
 block         3  NCSF(L) =       126  NCSF(G) =      3291  NCSF =     23926
 block         4  NCSF(L) =       128  NCSF(G) =      3401  NCSF =     23946
 block         5  NCSF(L) =       156  NCSF(G) =      4817  NCSF =     34879
 block         6  NCSF(L) =       152  NCSF(G) =      4959  NCSF =     34837
 block         7  NCSF(L) =       125  NCSF(G) =      5604  NCSF =     39901
 block         8  NCSF(L) =       125  NCSF(G) =      5789  NCSF =     39901
 block         9  NCSF(L) =        79  NCSF(G) =      5790  NCSF =     39380
 block        10  NCSF(L) =        77  NCSF(G) =      5964  NCSF =     39343
STOP Normal Exit ...
\end{Verbatim}
\noindent
Here {\tt NCSF(L)} gives the number of labeling CSFs and {\tt NCSF(G)} the number of labeling and generating CSFs.
{\tt NCSF} gives  
the number of CSFs in the expanded list, which should be exactly equal to those generated by the {\sc Grasp2018} program {\tt rcsfgenerate}. The full list of CSFs in {\sc Grasp2018} format is given in the file {\tt rcsf.out}.
\subsection{Perform spin-angular integrations}\label{sec:psain}
As the next step, we use {\tt rangular\_csfg\_mpi} to perform the spin-angular integrations and write the data to files in temporary directories as set by the {\tt disks} file. We start (if we have not done this before) by copying the output file {\tt rcsfg.out} from   {\tt rcsfggenerate\_csfg} to {\tt rcsfg.inp} and {\tt rlabel.out} to {\tt rlabel.inp}
\begin{Verbatim}[fontsize=\footnotesize]
>>cp rcsfg.out rcsfg.inp
>>cp rlabel.out rlabel.inp
\end{Verbatim}
We now run {\tt rangular\_csfg\_mpi} on 20 nodes
\begin{Verbatim}[fontsize=\footnotesize]
>>mpirun -np 20 rangular_csfg_mpi

 ====================================================
        RANGULAR_CSFG_MPI: Execution Begins ...
 ====================================================
    .....
    
 Full interaction?  (y/n)
>>y
    ....
    
 ====================================================
        RANGULAR_CSFG_MPI: Execution Finished ...
 ====================================================
\end{Verbatim}

\noindent
The user time for the run was \per{8.7s}, which should be compared with the time  \per{3m\,15s} for the {\tt rangular\_mpi} program of
{\sc Grasp2018}. The size of the spin-angular data is \per{0.96 GB}, which should be compared with \per{24 GB}  for {\tt rangular\_mpi}.
Both comparisons show the speed- and resource effectiveness of the new methodology based on CSFGs. It should be pointed out that the above computation loads (time and size of the spin-angular data) would be kept as constants for intended more large-scale calculations using larger orbital size, such as $n=8, 9, \ldots$, whereas they increase very fast for {\tt rangular\_mpi}, compare table \ref{tab_rangular}.

\subsection{Generate initial estimates of radial orbitals}
As a next step, we use {\tt rwfnestimate\_csfg} to generate initial estimates of radial orbitals. This step assumes that previous calculations have been done and that, in this case, the radial orbital file {\tt n6.w} is available. Additional files that are needed are {\tt isodata} and {\tt rcsfg.inp}. The input is as follows  
\begin{Verbatim}[fontsize=\footnotesize]
>>rwfnestimate_csfg

 RWFNESTIMATE_CSFG
 This program estimates radial wave functions
 for orbitals
 Input files: isodata, rcsfg.inp, optional rwfn file
 Output file: rwfn.inp
 Default settings ?
>>y
 Loading CSF file ... Header only
 There are/is           49  relativistic subshells;
 The following subshell radial wavefunctions remain to be estimated:
 1s 2s 2p- 2p 3s 3p- 3p 3d- 3d 4s 5s 6s 7s 4p- 5p- 6p- 7p- 4p 5p 6p 7p 4d- 5d-
 6d- 7d- 4d 5d 6d 7d 4f- 5f- 6f- 7f- 4f 5f 6f 7f 5g- 6g- 7g- 5g 6g 7g 6h- 7h-
 6h 7h 7i- 7i

 Read subshell radial wavefunctions. Choose one below
     1 -- GRASP92 File
     2 -- Thomas-Fermi
     3 -- Screened Hydrogenic
     4 -- Screened Hydrogenic [custom Z]
>>1
 Enter the file name (Null then "rwfn.out")
>>n6.w
 Enter the list of relativistic subshells:
>>*
 The following subshell radial wavefunctions remain to be estimated:
 7s 7p- 7p 7d- 7d 7f- 7f 7g- 7g 7h- 7h 7i- 7i

 Read subshell radial wavefunctions. Choose one below
     1 -- GRASP92 File
     2 -- Thomas-Fermi
     3 -- Screened Hydrogenic
     4 -- Screened Hydrogenic [custom Z]
>>3
 Enter the list of relativistic subshells:
>>*

    .....
    
 RWFNESTIMATE: Execution complete.
\end{Verbatim}
\subsection{Perform MCDHF calculation}
All data are now available, and we use {\tt rmcdhf\_csfg\_mpi} to perform the MCDHF calculation. The input is the same as for the
corresponding {\sc Grasp2018} program.
\begin{Verbatim}[fontsize=\footnotesize]
>>mpirun -np 20 rmcdhf_csfg_mpi

 ====================================================
        RMCDHF_CSFG_MPI: Execution Begins ...
 ====================================================
    ....

Default settings?  (y/n)
>>y

 Now perform rcsfsymexpand and fill NTYPE ...
    ....

 There are        42029  relativistic CSFs... load complete;

 Enter ASF serial numbers for each block
 Block         1   NCSF(G)s =      1200  NCSFs =      8579  id =    0+
>>1-5
 Block         2   NCSF(G)s =      1214  NCSFs =      8542  id =    0-
>>1-2
 Block         3   NCSF(G)s =      3291  NCSFs =     23926  id =    1+
>>1-3
 Block         4   NCSF(G)s =      3401  NCSFs =     23946  id =    1-
>>1-5
 Block         5   NCSF(G)s =      4817  NCSFs =     34879  id =    2+
>>1-7
 Block         6   NCSF(G)s =      4959  NCSFs =     34837  id =    2-
>>1-5
 Block         7   NCSF(G)s =      5604  NCSFs =     39901  id =    3+
>>1-2
 Block         8   NCSF(G)s =      5789  NCSFs =     39901  id =    3-
>>1-3
 Block         9   NCSF(G)s =      5790  NCSFs =     39380  id =    4+
>>1-2
 Block        10   NCSF(G)s =      5964  NCSFs =     39343  id =    4-
>>1
 level weights (1 equal;  5 standard;  9 user)
>>5
 Radial functions
 1s 2s 2p- 2p 3s 3p- 3p 3d- 3d 4s 5s 6s 7s 4p- 5p- 6p- 7p- 4p 5p 6p 7p 4d- 5d-
 6d- 7d- 4d 5d 6d 7d 4f- 5f- 6f- 7f- 4f 5f 6f 7f 5g- 6g- 7g- 5g 6g 7g 6h- 7h-
 6h 7h 7i- 7i
 Enter orbitals to be varied (Updating order)
>>7*
 Which of these are spectroscopic orbitals?
>>
 Enter the maximum number of SCF cycles:
>>100
    ....

 ====================================================
        RMCDHF_CSFG_MPI: Execution Finished ...
 ====================================================
\end{Verbatim}
The user time for the run was \per{7m\,18s}, which should be compared with  the  time \per{1h\,5m} for the {\tt rmcdhf\_mem\_mpi} program of
{\sc Grasp2018}.

\subsection{Saving the data}
We now run {\tt rsave\_csfg} to save the output data in accordance with the {\sc Grasp2018} and {\sc Graspg} naming conventions.
\begin{Verbatim}[fontsize=\footnotesize]
>>rsave_csfg n7
Created n7.w, n7.c, n7.m, n7.sum, n7.alog, n7.log, n7.g and n7.l
\end{Verbatim}
Here {\tt n7.w} contains the radial orbitals, {\tt n7.g} the CSF list in {\sc Graspg} format, {\tt n7.l} the information about the labeling orbitals, {\tt n7.c} the full list of CSFs obtained by de-exitation of the generating CSFs in  the
{\tt n7.g} file, {\tt n7.m} the mixing coefficients in relation to the CSFs in the {\tt n7.c} file, {\tt n7.sum} the summary of the
run giving the total number of CSFs, radial expectation values, energies and leading mixing coefficients, {\tt n7.alog} 
the log-file that mirrors the input to {\tt rangular\_csfg\_mpi}, and {\tt n7.log} 
the log-file that mirrors the input to {\tt rmcdhf\_csfg\_mpi}

\subsection{Generate list of CSFs in {\sc Graspg} format for the CI calculation}
We use {\tt rcsfggenerate\_csfg} to generate MR-SD expansion accounting for VV, CV, and CC electron correlation.
The $1s^2$ shell is closed in all CSFs. The generated list of CSFs in {\sc Graspg} format provides the input for the subsequent CI calculation.

\begin{Verbatim}[fontsize=\footnotesize]
>>rcsfggenerate_csfg

 RCSFGGENERATE_CSFG
 This program generates a list of CSFG CSFs

 Configurations should be entered in spectroscopic notation
 with occupation numbers and indications if orbitals are
 closed (c), inactive (i), active (*) or has a minimal
 occupation e.g. 1s(2,1)2s(2,*)
 Outputfiles: rcsfg.out, rlabel.out, rcsfggenerate.log

 Select core
        0: No core
        1: He (       1s(2)                  =  2 electrons)
        2: Ne ([He] + 2s(2)2p(6)             = 10 electrons)
        3: Ar ([Ne] + 3s(2)3p(6)             = 18 electrons)
        4: Kr ([Ar] + 3d(10)4s(2)4p(6)       = 36 electrons)
        5: Xe ([Kr] + 4d(10)5s(2)5p(6)       = 54 electrons)
        6: Rn ([Xe] + 4f(14)5d(10)6s(2)6p(6) = 86 electrons)
>>1

 Generation of CSFs in labeling space
 Entered orbitals define the labeling orbital set
 Enter list of (maximum 1000) configurations.
 End list with a blank line or an asterisk (*)

 Give configuration           1
>>2s(2,*)2p(6,*)3s(2,*)
 Give configuration           2
>>2s(2,*)2p(6,*)3p(2,*)
 Give configuration           3
>>2s(2,*)2p(6,*)3s(1,*)3d(1,*)
 Give configuration           4
>>2s(2,*)2p(6,*)3d(2,*)
 Give configuration           5
>>
 Give set of active orbitals, as defined by the highest principal quantum number
 in a comma delimited list in s,p,d etc order, e.g. 3s,3p,3d or 4s,4p,4d,4f or
 5s,5p,5d,5f,5g etc
>>3s,3p,3d
 Resulting 2*J-number? lower, higher (J=1 -> 2*J=2 etc.)
>0,8
 Number of excitations (if negative number e.g. -2, correlation
 orbitals will always be doubly occupied)
>>0
 Generate more lists ? (y/n)
>>y
 Generation of CSFs in labeling space and, if entered orbitals extend 
 the labeling orbital set, also generating CSFs in correlation space
 Orbitals extending the labeling orbital set appear symmetry ordered
 Enter list of (maximum 1000) configurations.
 End list with a blank line or an asterisk (*)

 Give configuration           1
>>2s(2,*)2p(6,*)3s(2,*)
 Give configuration           2
>>2s(2,*)2p(6,*)3p(2,*)
 Give configuration           3
>>2s(2,*)2p(6,*)3s(1,*)3d(1,*)
 Give configuration           4
>>2s(2,*)2p(6,*)3d(2,*)
 Give configuration           5
>>
 Give set of active orbitals, as defined by the highest principal quantum number
 per l-symmetry, in a comma delimited list in s,p,d etc order, e.g.  5s,4p,3d
>>7s,7p,7d,7f,7g,7h,7i
 Resulting 2*J-number? lower, higher (J=1 -> 2*J=2 etc.)
>>0,8
 Number of excitations (if negative number e.g. -2, correlation
 orbitals will always be doubly occupied)
>>2
 Generate more lists ? (y/n)
>>y
 Generation of CSFs in labeling space and, if entered orbitals extend 
 the labeling orbital set, also generating CSFs in correlation space
 Orbitals extending the labeling orbital set appear symmetry ordered
 Enter list of (maximum 1000) configurations. 
 End list with a blank line or an asterisk (*)

 Give configuration           1
>>2s(2,*)2p(6,*)3s(1,*)3p(1,*)
 Give configuration           2
>>2s(2,*)2p(6,*)3p(1,*)3d(1,*)
 Give configuration           3
>>
 Give set of active orbitals, as defined by the highest principal quantum number
 per l-symmetry, in a comma delimited list in s,p,d etc order, e.g.  5s,4p,3d
>>7s,7p,7d,7f,7g,7h,7i
 Resulting 2*J-number? lower, higher (J=1 -> 2*J=2 etc.)
>>0,8
 Number of excitations (if negative number e.g. -2, correlation
 orbitals will always be doubly occupied)
>>2
 Generate more lists ? (y/n)
>>n
    ....
    
 RCSFGGenerate_CSFG: Execution complete.
\end{Verbatim}
\noindent
The run results in a list with 603\,450 labeling and generating CSFs, which, if expanded, corresponds to 4\,060\,967 CSFs. 
\subsection{Perform CI calculation}
All files needed to perform the CI calculation are now available. We choose {\tt n7CI} as the name for the run, and 
start by copying the relevant files
\begin{Verbatim}[fontsize=\footnotesize]
>>cp rcsfg.out n7CI.g
>>cp rlabel.out n7CI.l
>>cp n7.w n7CI.w
\end{Verbatim}
\noindent
We use {\tt rci\_csfg\_mpi} to perform the calculation and distribute it over 20 nodes. 
However, before proceeding, we need to determine the amount of internal memory (RAM) available for the CI run on the computer at hand. This is done by issuing the command
\begin{Verbatim}[fontsize=\footnotesize]
>>cat /proc/meminfo
\end{Verbatim}
and on the computer for one of the authors (PJ) we have
\begin{Verbatim}[fontsize=\footnotesize]
MemTotal:       792264208 kB
MemFree:        505584752 kB
MemAvailable:   510877144 kB
    ....
\end{Verbatim}
Thus, in this particular case, the available memory per node is 510/20 \per{GB $\approx$ 25 GB}. The amount of memory per node determines if the computed Hamiltonian matrix can be kept in memory or has to be stored on disk during the diagonalization.
In this run, we choose to have no restrictions on the Breit interaction,
and we answer no to the questions about limitations. After answering no, we may input any limit for $n$ and $l$. 
We choose to input 100 to further emphasize that there are no restrictions. The input is as follows
\begin{Verbatim}[fontsize=\footnotesize]
>>mpirun -np 20 rci_csfg_mpi 

 ====================================================
        RCI_CSFG_MPI: Execution Begins ...
 ====================================================
    ....

 Default settings?
>>y
 Name of state:
>>n7CI
 Now perform rcsfsymexpand ...

 Include contribution of H (Transverse)?
>>y
 Modify all transverse photon frequencies?
>>y
 Enter the scale factor:
>>1.e-6
 Limit the Breit contributions by n?
>>n
 Discard Breit interaction for above n:
>>100
 Limit the Breit contributions by l?
>>n
 Discard Breit interaction for above l:
>>100
 Include H (Vacuum Polarisation)?
>>y
 Include H (Normal Mass Shift)?
>>n
 Include H (Specific Mass Shift)?
>>n
 Estimate self-energy?
>>y
 Largest n (NQEDMAX) quantum number for including self-energy for orbital
 It is also label the max n-value for spectroscopy orbitals
 In addition 1: it should be not larger than nmaxgen
 In addition 2: it should be less or equal 8
>>3
 Input MaxMemPerProcs (in GB) ...
>>25
 ....

 Enter ASF serial numbers for each block
 Block         1   NCSF(G)s =     17652  NCSFs =    120544  id =    0+
>>1-5
 Block         2   NCSF(G)s =     15144  NCSFs =    107191  id =    0-
>>1-2
 Block         3   NCSF(G)s =     49013  NCSFs =    338112  id =    1+
>>1-3
 Block         4   NCSF(G)s =     43024  NCSFs =    302971  id =    1-
>>1-5
 Block         5   NCSF(G)s =     74078  NCSFs =    502129  id =    2+
>>1-7
 Block         6   NCSF(G)s =     64398  NCSFs =    448156  id =    2-
>>1-5
 Block         7   NCSF(G)s =     88303  NCSFs =    587410  id =    3+
>>1-2
 Block         8   NCSF(G)s =     77179  NCSFs =    524879  id =    3-
>>1-3
 Block         9   NCSF(G)s =     93419  NCSFs =    597511  id =    4+
>>1-2
 Block        10   NCSF(G)s =     81240  NCSFs =    532064  id =    4-
>>1
    ....

 ====================================================
        RCI_CSFG_MPI: Execution Finished ...
 ====================================================
\end{Verbatim}
The following files are produced in the run: {\tt n7CI.c}, which contains the expanded CSF list in {\sc Grasp2018} format, {\tt n7CI.cm} which contains the mixing coefficients with respect to the CSFs in the {\tt n7CI.c} file, {\tt n7CI.csum}, which contains a summary of energies from the run, and {\tt n7CI.clog}, which is a log-file which mirrors  the input. The execution time is
\per{42m\,10s}. The time for the corresponding run using {\tt rci\_mpi} is \per{439m\,40s}.

As mentioned above, it is possible to restrict the  Breit interaction between CSFs with negligible or minor changes in transition energies and 
resulting wave \mbox{functions \cite{CSFG}}.
To consider the interaction only between CSFs built from orbitals with $l \le 3$ 
 ($s\mspace{1mu}p\mspace{1mu}d\mspace{1mu}f$ orbitals) we should respond
\begin{Verbatim}[fontsize=\footnotesize]
  Limit the Breit contributions by l?
>>y
 Discard Breit interaction for above l:
>>3
\end{Verbatim}
The execution time in this case is reduced to \per{27m\,45s}, and the mean relative change of the transition energies is of the order $4.7 \times 10^{-4}$ \%, which is completely negligible.
\subsection{Labeling in LSJ-coupling}
To be able to compare the transition energies from MCDHF or CI runs, the states need to be labeled in a way consistent to what is used in compilations, e.g., from NIST \cite{NIST}. To obtain labels in $LSJ$-coupling, we run {\tt jj2lsj\_csfg} as follows
\begin{Verbatim}[fontsize=\footnotesize]
>>jj2lsj_csfg

 jj2lsj_csfg: Transformation of ASFs from a jj-coupled CSF basis
              into an LS-coupled CSF basis  (Fortran 95 version)
              (C) Copyright by   G. Gaigalas and Ch. F. Fischer,
              (2024).
              Modifications using CSFGs      by C.Y. Chen (2023)
              Input files: name.c, name.(c)m, name.g, name.l
               (optional)  name.lsj.T
              Ouput files: name.lsj.lbl,
               (optional)  name.lsj.c, name.lsj.j,
                           name.uni.lsj.lbl, name.uni.lsj.sum,
                           name.lsj.T
              Optional file name.lsj.T is not available for
              jj2lsj_csfg

 Name of state
>>n7CI
 Now perform rcsfgexpand ...
    ....
 
 Loading Configuration Symmetry List File ...
 There are 49 relativistic subshells;
 There are 4060967 relativistic CSFs;
  ... load complete;

 Mixing coefficients from a CI calc.?
>>y
 Do you need a unique labeling? (y/n)
>>y
    nelec  =           12
    ncftot =      4060967
    nw     =           49
    nblock =           10

   block     ncf     nev    2j+1  parity
       1  120544       5       1       1
       2  107191       2       1      -1
       3  338112       3       3       1
       4  302971       5       3      -1
       5  502129       7       5       1
       6  448156       5       5      -1
       7  587410       2       7       1
       8  524879       3       7      -1
       9  597511       2       9       1
      10  532064       1       9      -1
 Default settings?  (y/n)
>>y
    ....

 jj2lsj_csfg: Execution complete.
\end{Verbatim}
\noindent
The {\tt jj2lsj\_csfg} produces the  {\tt n7CI.lsj.lbl} and {\tt n7CI.uni.lsj.lbl} files with labeling information.
The latter is used by the subsequent {\tt rlevels\_csfg} program.  
\subsection{Displaying the energy levels}\label{sec:displaye}
To display the energy levels, we use the {\tt rlevels\_csfg} program, which is a slight update of the corresponding {\sc Grasp} program.
\begin{Verbatim}[fontsize=\footnotesize]
>>rlevels_csfg n7CI.cm

 nblock =           10   ncftot =      4060967   nw =           49   nelec =           12

 Energy levels for ...
 Rydberg constant is   109737.31569
 Splitting is the energy difference with the lower neighbor
------------------------------------------------------------------------------------------
 No Pos  J Parity Energy Total    Levels     Splitting     Configuration
                      (a.u.)      (cm^-1)     (cm^-1)
------------------------------------------------------------------------------------------
  1  1   0  +   -1182.7219367        0.00        0.00  2s(2).2p(6).3s(2)_1S
  2  1   0  -   -1181.6559325   233960.89   233960.89  2s(2).2p(6).3s.3p_3P
  3  1   1  -   -1181.6293351   239798.34     5837.45  2p(6).3s.3p_3P
  4  1   2  -   -1181.5648710   253946.59    14148.24  2s(2).2p(6).3s.3p_3P
  5  2   1  -   -1181.1162782   352401.31    98454.73  2p(6).3s.3p_1P
  6  2   0  +   -1180.1932302   554986.94   202585.63  2p(6).3p(2)_3P
  7  1   2  +   -1180.1705927   559955.30     4968.35  2s(2).2p(6).3p(2)_1D
  8  1   1  +   -1180.1476150   564998.31     5043.01  2s(2).2p(6).3p(2)_3P
  9  2   2  +   -1180.0692436   582198.84    17200.53  2p(6).3p(2)_3P
 10  3   0  +   -1179.7124896   660497.30    78298.46  2p(6).3p(2)_1S
 11  2   1  +   -1179.6290841   678802.70    18305.40  2s(2).2p(6).3s.3d_3D
 12  3   2  +   -1179.6243821   679834.66     1031.96  2s(2).2p(6).3s.3d_3D
 13  1   3  +   -1179.6170137   681451.84     1617.18  2s(2).2p(6).3s.3d_3D
 14  4   2  +   -1179.2459175   762898.04    81446.20  2p(6).3s.3d_1D
 15  2   2  -   -1178.4913090   928515.46   165617.42  2s(2).2p(6).3p.3d_3F
 16  1   3  -   -1178.4462254   938410.17     9894.71  2s(2).2p(6).3p.3d_3F
 17  3   2  -   -1178.3991731   948736.96    10326.80  2s(2).2p(6).3p.3d_1D
 18  1   4  -   -1178.3937828   949919.98     1183.02  2s(2).2p(6).3p.3d_3F
 19  3   1  -   -1178.2423510   983155.42    33235.44  2s(2).2p(6).3p.3d_3D
 20  4   2  -   -1178.2393409   983816.06      660.64  2p(6).3p.3d_3P
 21  2   3  -   -1178.1876120   995169.26    11353.20  2s(2).2p(6).3p.3d_3D
 22  2   0  -   -1178.1827854   996228.57     1059.32  2s(2).2p(6).3p.3d_3P
 23  4   1  -   -1178.1812072   996574.94      346.36  2s(2).2p(6).3p.3d_3P
 24  5   2  -   -1178.1795441   996939.96      365.02  2s(2).2p(6).3p.3d_3D
 25  3   3  -   -1177.8753369  1063705.71    66765.75  2p(6).3p.3d_1F
 26  5   1  -   -1177.8182752  1076229.31    12523.60  2p(6).3p.3d_1P
 27  5   2  +   -1176.4776792  1370456.13   294226.82  2s(2).2p(6).3d(2)_3F
 28  2   3  +   -1176.4700401  1372132.71     1676.58  2s(2).2p(6).3d(2)_3F
 29  1   4  +   -1176.4607168  1374178.93     2046.22  2s(2).2p(6).3d(2)_3F
 30  6   2  +   -1176.3288764  1403114.55    28935.63  2s(2).2p(6).3d(2)_1D
 31  4   0  +   -1176.3167909  1405767.01     2652.46  2s(2).2p(6).3d(2)_3P
 32  3   1  +   -1176.3138687  1406408.37      641.36  2s(2).2p(6).3d(2)_3P
 33  2   4  +   -1176.3073349  1407842.38     1434.01  2s(2).2p(6).3d(2)_1G
 34  7   2  +   -1176.3068271  1407953.82      111.44  2p(6).3d(2)_3P
 35  5   0  +   -1175.9388990  1488704.70    80750.88  2s(2).2p(6).3d(2)_1S
------------------------------------------------------------------------------------------
\end{Verbatim}
\noindent
The computed energy levels compare well with the experimental ones reported by NIST \cite{NIST}.
\subsection{Condensing the CSF list}
The program {\tt rmixaccumulate\_csfg} can be used to condense an available list of labeling and generating CSFs, as well as to
predict which are the main contributing CSFs for a case based on a larger orbital set. We start by condensing the $n = 7$ CI expansion by retaining
all labeling and generating CSFs that contribute to 0.99999999 of the accumulated weight.

\begin{Verbatim}[fontsize=\footnotesize]
>>rmixaccumulate_csfg

 ***************************************************************************
 Welcome to program rmixaccumulate_csfg

 The program accumulates the squared weights of the CSFGs from smaller CI.
 The CSFGs in the output list can be sorted by mixing coefficents.
 A priori condensation expansions can be obtained by redefining the
 highest orbitals for each symmetry.

 Input files: <state>.(c)m, <state>.g, and <state>.l
 Output file: rcsfg.out, rcsfg.rem, rcsf.out,
              rcsfg.nmax(optional)

  Modified from rmixaccumulate.f90 written by J. Ekman & P. Jonsson Feb 2016
                                CSFG-version written by Yanting Li, Apr 2022
                                          Modify by  Chongyang Chen     2023
 ***************************************************************************

 Give name of the state:
>>n7CI
 Expansion coefficients resulting from CI calculation (y/n)?
>>y
 Fraction of total wave function [0-1] to be included in reduced list:
>>0.99999999
 Defining CSFs in output file sorted by mixing coefficients (y/n)?
>>y

 Now perform rcsfsymexpand ...
    ....

 Number of CSFGs written to rcsfg.out
         block    NCSF(G)
           1       11033
           2        8933
           3       30835
           4       29573
           5       52577
           6       44918
           7       49860
           8       48304
           9       48890
          10       35303
 Redefine the CSFGs (y/n)?
>>n

 Now perform rcsfgexpand ...
 block         1  NCSF(L) =       516  NCSF(G) =     11033  NCSF =     81193
 block         2  NCSF(L) =       421  NCSF(G) =      8933  NCSF =     68763
 block         3  NCSF(L) =      1239  NCSF(G) =     30835  NCSF =    226089
 block         4  NCSF(L) =      1115  NCSF(G) =     29573  NCSF =    220455
 block         5  NCSF(L) =      1649  NCSF(G) =     52577  NCSF =    380088
 block         6  NCSF(L) =      1421  NCSF(G) =     44918  NCSF =    332407
 block         7  NCSF(L) =      1489  NCSF(G) =     49860  NCSF =    357471
 block         8  NCSF(L) =      1321  NCSF(G) =     48304  NCSF =    353721
 block         9  NCSF(L) =      1129  NCSF(G) =     48890  NCSF =    338788
 block        10  NCSF(L) =       967  NCSF(G) =     35303  NCSF =    248096
STOP Normal Exit
\end{Verbatim}
\noindent
The program outputs three files: {\tt rcsfg.out}, which contains the labeling and generating CSFs of the condensed list, {\tt rcsfg.rem}, which contains the removed generating CSFs, and {\tt rcsf.out}, which contains the CSFs obtained by expanding the condensed list by applying the de-excitation rules. 
The number of CSFs, {\tt NCSF}, in the {\tt rcsf.out} file is given in the rightmost column following the last call to  {\tt rcsfgexpand}.
The time for running {\tt rci\_csfg\_mpi} based on the condensed list is \per{26m\,7s} compared with \per{42m\,10s} based on the full list.
The mean relative difference in transition energies based on the condensed
and full list is $0.00001$ \%, 
%\gg{!!{\bf Comment: I would like suggest to write as $0.00001$ \% }} 
i.e., virtually zero. Applying restrictions to the Breit interaction, amounting
to including the interaction only between CSFs built from the labeling orbitals, further reduce the execution time to \per{10m\,23s}, again with negligible difference in transition energies.  
\subsection{Predicting the most important CSFs for an extended orbital set}
To predict which are the most important CSFs for the symmetry ordered orbital set extended to $n=8$ we use {\tt rmixaccumulate\_csf} in the following way.
\begin{Verbatim}[fontsize=\footnotesize]
>>rmixaccumulate_csfg

 ***************************************************************************
 Welcome to program rmixaccumulate_csfg

 The program accumulates the squared weights of the CSFGs from smaller CI.
 The CSFGs in the output list can be sorted by mixing coefficents.
 A priori condensation expansions can be obtained by redefining the
 highest orbitals for each symmetry.

 Input files: <state>.(c)m, <state>.g, and <state>.l
 Output file: rcsfg.out, rcsfg.rem, rcsf.out,
              rcsfg.nmax(optional)

  Modified from rmixaccumulate.f90 written by J. Ekman & P. Jonsson Feb 2016
                                CSFG-version written by Yanting Li, Apr 2022
                                          Modify by  Chongyang Chen     2023
 ***************************************************************************

 Give name of the state:
>>n7CI
 Expansion coefficients resulting from CI calculation (y/n)?
>>y
 Fraction of total wave function [0-1] to be included in reduced list:
>>0.99999999
 Defining CSFs in output file sorted by mixing coefficients (y/n)?
>>y

 Now perform rcsfsymexpand ...
    ....
    
 Number of CSFGs written to rcsfg.out
         block    NCSF(G)
           1       11033
           2        8933
           3       30835
           4       29573
           5       52577
           6       44918
           7       49860
           8       48304
           9       48890
          10       35303
 Redefine the CSFGs (y/n)?
>>y
   2s   2p-  2p   3s   3p-  3p   3d-  3d   4s   5s   6s   7s   4p-  5p-  6p-  7p- 
   4p   5p   6p   7p   4d-  5d-  6d-  7d-  4d   5d   6d   7d   4f-  5f-  6f-  7f-  
   4f   5f   6f   7f   5g-  6g-  7g-  5g   6g   7g   6h-  7h-  6h   7h   7i-  7i
 Give set of active orbitals, as defined by the highest principalquantum number
 per l-symmetry, in a comma delimited list in s,p,d etc order,e.g. 5s,4p,3d
>>8s,8p,8d,8f,8g,8h,8i
           9           1
 Redefined symmetry-ordered set:
   2s   2p-  2p   3s   3p-  3p   3d-  3d   4s   5s   6s   7s   8s   4p-  5p-  6p- 
   7p-  8p-  4p   5p   6p   7p   8p   4d-  5d-  6d-  7d-  8d-  4d   5d   6d   7d   
   8d   4f-  5f-  6f-  7f-  8f-  4f   5f   6f   7f   8f   5g-  6g-  7g-  8g-  5g   
   6g   7g   8g   6h-  7h-  8h-  6h   7h   8h   7i-  8i-  7i   8i

 Now perform rcsfsymexpand ...
 block         1  NCSF(L) =       516  NCSF(G) =     11033  NCSF =    127333
 block         2  NCSF(L) =       421  NCSF(G) =      8933  NCSF =    108790
 block         3  NCSF(L) =      1239  NCSF(G) =     30835  NCSF =    359678
 block         4  NCSF(L) =      1115  NCSF(G) =     29573  NCSF =    355261
 block         5  NCSF(L) =      1649  NCSF(G) =     52577  NCSF =    616250
 block         6  NCSF(L) =      1421  NCSF(G) =     44918  NCSF =    542329
 block         7  NCSF(L) =      1489  NCSF(G) =     49860  NCSF =    580472
 block         8  NCSF(L) =      1321  NCSF(G) =     48304  NCSF =    580614
 block         9  NCSF(L) =      1129  NCSF(G) =     48890  NCSF =    557589
 block        10  NCSF(L) =       967  NCSF(G) =     35303  NCSF =    408039
STOP Normal Exit
\end{Verbatim}
\noindent
Now the program outputs four files: {\tt rcsfg.out}, which contains the labeling and generating CSFs of the condensed list, {\tt rcsfg.rem}, which contains the removed generating CSFs, {\tt rcsfg.nmax} which contains the labeling and generating CSFs of the condensed list, but with the redefined upper limits 
of the orbitals in the symmetry ordered set, and {\tt rcsf.out}, which contains the CSFs obtained by expanding the {\tt rcsfg.nmax} list by applying the de-excitation rules. 
The number of CSFs, {\tt NCSF}, in the {\tt rcsf.out} file is given in the rightmost column following the last call to  {\tt rcsfgexpand}.
The condensed list for $n = 8$ contains approximately a factor 2 less CSFs compared to the corresponding uncondensed list for $n=8$.
\subsection{Perform CI calculation with part of correlation treated perturbatively}
As discussed in section \ref{sec:PT}, {\sc Graspg} has options to include parts of the interaction
in the Hamiltonian matrix perturbatively.
As an illustration we partition the CSFs in {\tt n7CI.g} accounting for VV, CV, and CC correlation into 
a zero-order space, $P$, accounting for VV and CV correlation and a first-order space, $Q$, accounting for CC correlation, see  \cite{Stefan} for a similar
case.
The zero-order space is available in {\tt n7.g}, and we run the
{\tt rcsfgzerofirst\_csfg} program as follows
\begin{Verbatim}[fontsize=\footnotesize]
>>rcsfgzerofirst_csfg

 RCSFGzerofirst_CSFG:
             Takes a list of CSFs and partitions each symmetry
             block into a zero- and first-order CSF space from
             a zero-order list.
             (C)   Copyright by G. Gaigalas and Ch. F. Fischer
             (Fortran 95 version)                NIST  (2021).

             CSFG version, Chongyang Chen,       Fudan (2023).

             Input files:    list with CSF(G)s to be partitioned
                             list with CSF(G)s defining
                                  the zero-order space
             Input files:    together with their labeling files

             Output file:    rcsfg.out
             Output file:    icut

 Give the full name of the list that contains the zero-order space
>>n7.g
 Give the full name of the list that should be partitioned
n7CI.g
 Loading Configuration Symmetry List File ...
 Checking labeling orbital files by diff n7.l n7CI.l
 There are 49 relativistic subshells;
 Block            Zero-order Space   Full Space
 block         1  NUMZ=      1200  NUMF=     17652  NFOUND=      1200
 block         2  NUMZ=      1214  NUMF=     15144  NFOUND=      1214
 block         3  NUMZ=      3291  NUMF=     49013  NFOUND=      3291
 block         4  NUMZ=      3401  NUMF=     43024  NFOUND=      3401
 block         5  NUMZ=      4817  NUMF=     74078  NFOUND=      4817
 block         6  NUMZ=      4959  NUMF=     64398  NFOUND=      4959
 block         7  NUMZ=      5604  NUMF=     88303  NFOUND=      5604
 block         8  NUMZ=      5789  NUMF=     77179  NFOUND=      5789
 block         9  NUMZ=      5790  NUMF=     93419  NFOUND=      5790
 block        10  NUMZ=      5964  NUMF=     81240  NFOUND=      5964

 Wall time:
       17 seconds

 Finish Date and Time:
   Date (Yr/Mon/Day): 2024/08/15
   Time (Hr/Min/Sec): 13/39/18.033
   Zone: +0200

 RCSFGzerofirst_CSFG: Execution complete.
\end{Verbatim}
The  output file {\tt rcsfg.out} contains the partitioned CSF list in {\sc Graspg} format.
The file {\tt icut} contains the number of CSFs in the zero-order space. 
We now copy the output file {\tt rcsfg.out} to {\tt n7CI\_ZF.g},
the {\tt n7.l} to {\tt n7CI\_ZF.l} and {\tt n7.w} to {\tt n7CI\_ZF.w}
\begin{Verbatim}[fontsize=\footnotesize]
>>cp rcsfg.out n7CI_ZF.g
>>cp n7.w n7CI_ZF.w
>>cp n7.l n7CI_ZF.l
\end{Verbatim}
The input to the CI program is
\begin{Verbatim}[fontsize=\footnotesize]
>>mpirun -np 20 rci_csfg_mpi

 ====================================================
        RCI_CSFG_MPI: Execution Begins ...
 ====================================================
    ....

 Default settings?
>>n
 Name of state:
>>n7CI_ZF

 Now perform rcsfsymexpand ...
    ...
    
 Restarting RCI90 ?
>>n
 Revise the physical speed of light (   137.03599913900001       in a.u.) ?
>>n
 Treat contributions of some CSFs as first-order perturbations?
>>y
 Include symmetry-ordered-diagonal block? (y/n)
>>n

 Enter iccut for each block
 Block          1   NCSF(G) =      17652  id =    0+
>>1200
 Block          2   NCSF(G) =      15144  id =    0-
>>1214
 Block          3   NCSF(G) =      49013  id =    1+
>>3291
 Block          4   NCSF(G) =      43024  id =    1-
>>3401
 Block          5   NCSF(G) =      74078  id =    2+
>>4817
 Block          6   NCSF(G) =      64398  id =    2-
>>4959
 Block          7   NCSF(G) =      88303  id =    3+
>>5604
 Block          8   NCSF(G) =      77179  id =    3-
>>5789
 Block          9   NCSF(G) =      93419  id =    4+
>>5790
 Block         10   NCSF(G) =      81240  id =    4-
>>5964
 Include contribution of H (Transverse)?
>>y
 Modify all transverse photon frequencies?
>>y
 Enter the scale factor:
>>1.e-6
 Limit the Breit contributions by n?
>>n
 Discard Breit interaction for above n:
>>100
 Limit the Breit contributions by l?
>>n
 Discard Breit interaction for above l:
>>100
 Include H (Vacuum Polarisation)?
>>y
 Include H (Normal Mass Shift)?
>>n
 Include H (Specific Mass Shift)?
>>n
 Estimate self-energy?
>>y
 Largest n (NQEDMAX) quantum number for including self-energy for orbital
 It is also label the max n-value for spectroscopy orbitals
 In addition 1: it should be not larger than nmaxgen
 In addition 2: it should be less or equal 8
>>3
 Input MaxMemPerProcs (in GB) ...
>>20
   ....
 
 Enter ASF serial numbers for each block
 Block          1   NCSF(G)s =     17652  NCSFs =    120544  id =    0+
>>1-5
 Block          2   NCSF(G)s =     15144  NCSFs =    107191  id =    0-
>>1-2
 Block          3   NCSF(G)s =     49013  NCSFs =    338112  id =    1+
>>1-3
 Block          4   NCSF(G)s =     43024  NCSFs =    302971  id =    1-
>>1-5
 Block          5   NCSF(G)s =     74078  NCSFs =    502129  id =    2+
>>1-7
 Block          6   NCSF(G)s =     64398  NCSFs =    448156  id =    2-
>>1-5
 Block          7   NCSF(G)s =     88303  NCSFs =    587410  id =    3+
>>1-2
 Block          8   NCSF(G)s =     77179  NCSFs =    524879  id =    3-
>>1-3
 Block          9   NCSF(G)s =     93419  NCSFs =    597511  id =    4+
>>1-2
 Block         10   NCSF(G)s =     81240  NCSFs =    532064  id =    4-
>>1
    ....
    
 ====================================================
        RCI_CSFG_MPI: Execution Finished ...
 ====================================================
 \end{Verbatim}
 \noindent
 The time for the run is \per{8m\,27s} and the mean relative energy difference with the full calculation is 0.058 \%. 
\subsection{Large CI calculations by $Jp$-block}
The preferred way to perform CI calculations for large cases is by $Jp$-block. Use {\tt rcsfgblocksplit\_csfg} to split the CSF list in {\sc Graspg} format to several lists, one for each block. For each block, use {\tt rci\_block\_csfg\_mpi} to perform the computation of Hamiltonian matrix distributed over many nodes. The latter scales almost linearly with the number of nodes. Use {\tt rdistHmatrix\_csfg}  to redistribute 
the resulting {\tt rci.res} files to fewer nodes. Then, finally, use {\tt rci\_block\_csfg\_mpi} in restart mode to perform the diagonalization, see \cite{GRASPCPU}. Note that a second temporary directory must be available
to hold the redistributed {\tt rci.res} files. Good advice is to give it the name {\tt tmpblock}.
We start by splitting the {\tt n7CI.g} file into blocks
\begin{Verbatim}[fontsize=\footnotesize]
>>rcsfgblocksplit_csfg

 RCSFGBLOCKSPLIT_CSFG
 Splits a CSF file name.g into corresponding files
 for each parity and J block
 Input file: name.g, name.l
 Output files: name_even1.g, name_even1.l
 name_odd1.g, name:odd1.l etc

 Name of the state
>>n7CI

 Each of the blocks must be built from the same orbital set
 This may not be true for MR expansions, but is normally true
 for SD-MR expansions
 Is the above condition fullfilled? (y,n)
>>y

 nblock              10
 nblockodd            5
 nblockeven           5


 Exit status of the name.inf file copying for block        1 was           0
 Exit status of the name.inf file copying for block        2 was           0
 Exit status of the name.inf file copying for block        3 was           0
 Exit status of the name.inf file copying for block        4 was           0
 Exit status of the name.inf file copying for block        5 was           0
 Exit status of the name.inf file copying for block        6 was           0
 Exit status of the name.inf file copying for block        7 was           0
 Exit status of the name.inf file copying for block        8 was           0
 Exit status of the name.inf file copying for block        9 was           0
 Exit status of the name.inf file copying for block       10 was           0
\end{Verbatim}
\noindent
The resulting files are {\tt n7CI\_even1.g}, {\tt n7CI\_even1.l}, {\tt n7CI\_odd1.g}, {\tt n7CI\_odd1.l} etc.
As a demonstration will run the CI program for the even parity $J=0$ states ({\tt even1} block)
and start by copying the radial wave function file
\begin{Verbatim}[fontsize=\footnotesize]
>>cp n7CI.w n7CI_even1.w
\end{Verbatim}
\noindent
Since we just want to compute the Hamiltonian matrix, we should not require any states, but just press return when asked about the ASF serial numbers for each block, and the input is as follows
\begin{Verbatim}[fontsize=\footnotesize]
>>mpirun -np 40 rci_block_csfg_mpi

 ====================================================
        RCI_BLOCK_CSFG_MPI: Execution Begins ...
 ====================================================
    ....
    
 Default settings?
>>y
 Name of state:
>>n7CI_even1
 Calling rcsfsymexpand .....
    ...

 Include contribution of H (Transverse)?
>>y
 Modify all transverse photon frequencies?
>>y
 Enter the scale factor:
>>1.e-6
 Limit the Breit contributions by n?
>>n
 Discard Breit interaction for above n:
>>100
 Limit the Breit contributions by l?
>>n
 Discard Breit interaction for above l:
>>100
 Include H (Vacuum Polarisation)?
>>y
 Include H (Normal Mass Shift)?
>>n
 Include H (Specific Mass Shift)?
>>n
 Estimate self-energy?
>>y
 Largest n (NQEDMAX) quantum number for including self-energy for orbital
 It is also label the max n-value for spectroscopy orbitals
 In addition 1: it should be not larger than nmaxgen
 In addition 2: it should be less or equal 8
>>3
 Input MaxMemPerProcs (in GB) ...
>>12

 Enter ASF serial numbers for each block
 Block          1   NCSF(G)s =     17652  NCSFs =    120544  id =    0+
>>

 No state is selected, just build H-matrix ...
    ....

 ====================================================
        RCI_BLOCK_CSFG_MPI: Execution Finished ...
 ====================================================
\end{Verbatim}
\noindent
To redistribute the {\tt rci.res} files, holding the Hamiltonian, to a structure consistent with fewer MPI process, 10 in this case, we
run the {\tt rdistHmatrix\_csfg} program in the following way (the paths have to be adapted by the user)

\begin{Verbatim}[fontsize=\footnotesize]
>>rdistHmatrix_csfg

Input the (absolute path) directory of the source HM files:
>>/tank/tspejo/graspruns/testgraspg/FeXV/tmp
Input the (absolute path) directory of the target HM files:
>>/tank/tspejo/graspruns/testgraspg/FeXV/tmpblock
Input the number of source HM files:
>>40
Input the number of target HM files:
>>10
 Write the head lines of rciXXX.res ...000
 Write the head lines of rciXXX.res ...001
 Write the head lines of rciXXX.res ...002
 Write the head lines of rciXXX.res ...003
 Write the head lines of rciXXX.res ...004
 Write the head lines of rciXXX.res ...005
 Write the head lines of rciXXX.res ...006
 Write the head lines of rciXXX.res ...007
 Write the head lines of rciXXX.res ...008
 Write the head lines of rciXXX.res ...009
STOP Normal Exit ...
\end{Verbatim}
\noindent
Note very carefully that before running the {\tt rci\_block\_csfg\_mpi} program in restart mode on 10 nodes to diagonalize the Hamiltonian matrix, the {\tt disks} file needs to be updated
to give the path to the directory where the {\tt rci.res} files now reside after redistribution. Assuming that this has been done, we issue the commands
\begin{Verbatim}[fontsize=\footnotesize]
>>mpirun -np 10 rci_block_csfg_mpi

 ====================================================
        RCI_BLOCK_CSFG_MPI: Execution Begins ...
 ====================================================

 Default settings?
>>n
 Name of state:
>>n7CI_even1
 Calling rcsfsymexpand .....
    ...

 Restarting RCI90 ?
>>y
 Estimate contributions from the self-energy?
>>y

Reduce the accuracy of eigenvalue to speed up the calculation, 
in case of the calculation performed for rmixacculate_csfg.
If yes, CRITE of maneigmpi.f90 would be set as 1.0d-12.
>>n
 Input MaxMemPerProcs (in GB) ...
>>50

 Enter ASF serial numbers for each block
 Block        1   NCSF(G)s =   17652  NCSFs =  120544  id =    0+
>>1-5
    ....
    
 ====================================================
        RCI_BLOCK_CSFG_MPI: Execution Finished ...
 ====================================================
\end{Verbatim}
\noindent
Comparing with the energies in section \ref{sec:displaye} we see that we get identical energies for the even parity $J= 0$ states.
\section{Scaling of execution time and memory  requirement }\label{sec:scaling} % export MPI_TMP="/home/per/tmp_mpi"                 
The {\sc Graspg} program package becomes relatively more efficient compared to {\sc Grasp2018} the larger the calculations are, as pointed out in \cite{CSFG,GRASPCPU}.
To quantify this, we consider a list of  $M'$ 
labeling and generating CSFs in {\sc Graspg} format spanning $M$ CSFs, corresponding to a reduction ratio $R = M/M'$. 
The reduction ratio $R$ depends on the rules to form the CSFs as well as the size of the orbital set.
The reduction ratios can be computed based on the output information from {\tt rcsfgexpand\_csfg}.
\subsection{Spin-angular integrations and MCDHF calculations}
In table \ref{tab:R} we display the block averaged (there are ten $Jp$-blocks) reduction ratio for the VV and CV MCDHF calculations in section \ref{sec:FEXV} as a function of the 
increasing orbital set.\clearpage

\begin{table}[h!]
\caption{Block averaged labeling and generating CSFs, $M'$, block averaged CSFs in the expanded list, $M$, and the reduction ratio $R = M/M'$ as functions of the increasing orbital set 
for the VV and CV MCDHF calculations reported in section \ref{sec:FEXV}. Orbitals given in non-relativistic notation.}
\centering
\label{tab:R}
    \begin{tabular}{lrrrr} \hline
        orbital set & \gg{$M'$} & \gg{$M$} & $R$ & $R^2$\\ \hline
        $\{5s,5p,5d,5f,5g\}$       & 2~431 &     5~606  & 2.3  & 5.3\\ 
        $\{6s,6p,6d,6f,6g,6h\}$    & 3~322 &    14~355  & 4.3  & 18.7\\
        $\{7s,7p,7d,7f,7g,7h,7i\}$ & 4~203 &    29~323  & 7.0  & 48.7 \\
        $\{8s,8p,8d,8f,8g,8h,8i\}$ & 4~313 &    49~915  & 11.6 & 133.9\\
        $\{9s,9p,9d,9f,9g,9h,9i\}$ & 4~313 &  76~130 & 17.6 & 311.5\\ \hline
    \end{tabular}
\end{table}

The {\tt rangular\_mpi} program in {\sc Grasp2018} performs $M(M+1)/2$ spin-angular integrations to
resolve the Dirac-Coulomb Hamiltonian matrix into spin-angular coefficients and integrals and writes the results to file, see  
\ref{sec:mcp} and \ref{sec:saCSFG}. Disregarding the fact that 
more than one spin-angular integration has to be performed for some combinations of generating CSFs, see \ref{sec:saCSFG},
the {\tt rangular\_csfg\_mpi} program in {\sc Graspg} performs only $M'(M'+1)/2$ spin-angular integrations, with expected speed-up 
and file size reduction factors scaling of roughly $R^2$.  The execution times and file sizes for MPI calculations based on 20 processes
for the {\tt rangular\_mpi} and {\tt rangular\_csfg\_mpi} programs are shown in table \ref{tab_rangular} along with the actual speed-up and file size reduction factors. The expected speed-up 
and file size reduction factors would eventually approach to $R^2$ when the number of symmetry-ordered orbitals are enlarged enough, because the computation loads for {\tt rangular\_csfg\_mpi} would never increase at a specific orbital set, as pointed out in section \ref{sec:psain}. This feature should be of great interest for many-body perturbation theory, where very large basis sets are generally employed.

\begin{table}[h!]
\caption{  User time of {\tt rangular\_mpi} and {\tt rangular\_csfg\_mpi}, as well as the size of the spin-angular data and their ratio as functions of increasing orbital set. }
\centering
\label{tab_rangular}
\begin{tabular}{lrrrrrr} \hline
\multicolumn{1}{c}{\multirow{2}{2cm}{Orbital Set}} & \multicolumn{2}{c}{\tt rangular\_mpi} & \multicolumn{2}{c}{\tt rangular\_csfg\_mpi} & \multicolumn{2}{c}{ratios} \\
 \cline{2-3} \cline{4-5} \cline{6-7}
& Time & Size & Time & Size & Time & Size \\
   \hline
$\{5s,5p,5d,5f,5g\}$ & 10s & 1.3~GB & 3.2s & 0.41~GB & 3.1 & 3.1\\
$\{6s,6p,6d,6f,6g,6h\}$ & 57s & 6.0~GB & 5.5s & 0.66~GB & 10 & 9.0\\
$\{7s,7p,7d,7f,7g,7h,7i\}$ &  3m~15s & 24~GB & 8.7s & 0.96~GB & 22 & 25\\
$\{8s,8p,8d,8f,8g,8h,8i\}$ &  10m~6s & 44~GB & 11s & 1.0~GB & 57 & 44\\
$\{9s,9p,9d,9f,9g,9h,9i\}$ & 24m~2s & 98~GB & 13s & 1.0~GB & 114 & 98\\
        \hline
\end{tabular}
\end{table}

\clearpage
Turning to the MCDHF calculations we are interested in the speed-up and memory (RAM) reduction factors. The speed-up for the 
 {\tt rmcdhf\_csfg\_mpi}
program compared to {\tt rmcdhf\_mpi} is mainly due to the redesigned and efficient constructions of direct- and 
exchange potentials, as well as Lagrange multipliers, and additional parallelization of the
diagonalization procedure, as  reported in section \ref{sec:mcdhfCSFG}. The memory reduction factor is due to the use of CSF lists in {\sc Graspg} format. Table \ref{tab_rmcdhf} reports the averaged CPU time for each iteration in the  {\tt rmcdhf\_mpi} and {\tt rmcdhf\_csfg\_mpi} calculations, as well as the memory usage as functions of increasing orbital set. The reductions in memory implies that the spin-angular data for larger calculations can be buffed in memory, removing the need for repeated and time consuming reading of spin-angular data from file. 

\begin{table}[h!]
\caption{The averaged CPU time for each loop taken in the  {\tt rmcdhf\_csfg\_mpi} and {\tt rmcdhf\_mpi} calculation, as well as the memory usage as functions of increasing orbital set.}
\centering
\label{tab_rmcdhf}
\begin{tabular}{lrrrrrr} \hline
\multicolumn{1}{c}{\multirow{2}{2cm}{Orbital Set}} & \multicolumn{2}{c}{\tt rmcdhf\_mpi} & \multicolumn{2}{c}{\tt rmcdhf\_csfg\_mpi} & \multicolumn{2}{c}{ratios} \\
 \cline{2-3} \cline{4-5} \cline{6-7}
& Time & \gg{RAM} & Time & \gg{RAM} & Time & \gg{RAM} \\
   \hline
%$\{4s,4p,4d,4f\}$ & 11s & 0.76~GB & 0.32s & 0.57~GB  & 34 & 1.3 \\
$\{5s,5p,5d,5f,5g\}$ & 7.9s & 2.0~GB & 1.8s & 0.89~GB & 4.3 & 2.3\\
$\{6s,6p,6d,6f,6g,6h\}$ & 27s & 5.7~GB & 7.1s & 1.7~GB & 3.8 & 3.4\\
$\{7s,7p,7d,7f,7g,7h,7i\}$ &  3m~32s & 18~GB & 24s & 3.4~GB & 8.8 & 5.3\\
$\{8s,8p,8d,8f,8g,8h,8i\}$ &  20m~3s & 46~GB & 1m~6s & 5.8~GB & 24 & 8.0\\
$\{9s,9p,9d,9f,9g,9h,9i\}$ & 1h~23m & 98~GB & 2m~24s & 10~GB & 34 & 9.6\\
        \hline
\end{tabular}
\end{table}

\subsection{CI calculations}
In table \ref{tab:R2} we display the block averaged (there are ten $Jp$-blocks) reduction ratio for the VV, CV and CV CI calculations in section \ref{sec:FEXV} as a function of the 
increasing orbital set.\clearpage

\begin{table}[h!]
\caption{Block averaged labeling and generating CSFs, $M'$, block averaged CSFs in the expanded list, $M$, and the reduction ratio $R = M/M'$ as functions of the increasing orbital set 
for the VV, CV and CC CI calculations reported in section \ref{sec:FEXV}. Orbitals given in non-relativistic notation.}
\centering
\label{tab:R2}
    \begin{tabular}{lrrrr} \hline
        orbital set & $M'$ & $M$ & $R$ & $R^2$\\ \hline
        $\{5s,5p,5d,5f,5g\}$       & 33\,114 &     74~766  & 2.2  & 4.8\\ 
        $\{6s,6p,6d,6f,6g,6h\}$    & 46\,580 &    195~288  & 4.2  & 17.6\\
        $\{7s,7p,7d,7f,7g,7h,7i\}$ & 60\,345 &    406~097  & 6.7  & 44.9 \\
        $\{8s,8p,8d,8f,8g,8h,8i\}$ & 62\,098 &    697~172  & 11.2 & 125.4\\
        $\{9s,9p,9d,9f,9g,9h,9i\}$ & 62\,098 &  1\,068\,514 & 17.2 & 295.8\\ \hline
    \end{tabular}
\end{table}

To compute that Hamiltonian matrix the {\tt rci\_mpi} program of {\sc Grasp2018} performs $M(M+1)/2$ spin-angular integrations.
The matrix elements then follow by multiplying the spin-angular coefficients with the corresponding radial integrals and then summing the contributions.
The {\tt rci\_csfg\_mpi} program in {\sc Graspg} performs only $M'(M'+1)/2$ spin-angular integrations. 
The matrix elements between the CSFs generated by the CSFGs follow by multiplying the spin-angular coefficients with the
radial integrals as obtained by the de-excitation rule and then summing the contributions. Although fast, the time for repeatedly applying the de-excitation rule, multiplying with the spin-angular coefficients and summing to  form all the matrix elements between the CSFs generated by the CSFGs can not be neglected in comparison with the time for the spin-angular integration between the defining CSFs. Thus we can never
expect to attain the $R^2$ scaling. In addition, the time for the diagonalization of the Hamiltonian matrix can not be neglected, especially for cases targeting many eigenvalue pairs.  The execution times and file sizes for MPI calculations based on 20 processes
for the {\tt rci\_mpi} and {\tt rci\_csfg\_mpi} programs are shown in table \ref{tab_rci} along with the actual speed-up.
Restricting the Breit interaction to CSFs built on $spdf$ orbitals, the execution time for {\tt rci\_csfg\_mpi} calculation for $n = 9$ goes down to \per{77m\,52s}, corresponding to
a speed-up factor of 46. Further applying, {\em a priori} condensation to an accumulated fraction of 0.99999999 based on the results for $n = 7$
brings the execution time down to \per{17m\,9s}, which in total gives a speed-up ratio of 208 with a negligible change in transition energies.

\begin{table}[h!]
\caption{ CPU time of {\tt rci\_csfg\_mpi} and {\tt rci\_mpi}, and their ratio as functions of increasing orbital set. 
The second last row gives the execution time with the Breit interaction restricted to CSFs built on $spdf$ orbitals  whereas the last row
shows the execution time with the Breit interaction restricted to CSFs built on $spdf$ orbitals based on an {\em a priori} condensed list of CSFs.}
\begin{center}
\label{tab_rci}
\begin{tabular}{lrrrrrr} \hline
Orbital Set & {\tt rci\_mpi} & {\tt rci\_csfg\_mpi} & ratio \\ \hline
$\{4s,4p,4d,4f\}$ &  1m~43s  & 1m~20s   &  1.3 \\
$\{5s,5p,5d,5f,5g\}$ &  15m~52s & 5m~41s &  2.8 \\
$\{6s,6p,6d,6f,6g,6h\}$ & 1h~37m  &  16m~13s & 6.0 \\
$\{7s,7p,7d,7f,7g,7h,7i\}$ &  7h~19m  & 42m~10s  & 10.4 \\
$\{8s,8p,8d,8f,8g,8h,8i\}$ & 23h~18m &  1h~29m  &  15.7 \\
$\{9s,9p,9d,9f,9g,9h,9i\}$ & 2d~11h  &  2h~50m  & 20.9 \\ \hline
$\{9s,9p,9d,9f,9g,9h,9i\}^*$ &         &  1h~17m  & 46.0 \\
$\{9s,9p,9d,9f,9g,9h,9i\}^{**}$ &         &  17m~9s  & 208 \\
        \hline
\end{tabular}\\
\end{center}
\footnotesize{$^*$ Breit interaction restricted to CSFs built on $spdf$ orbitals. \\
$^{**}$ Restricted Breit and {\em a priori} condensation.}
\end{table}

\clearpage
\section{Acknowledgment}
The paper is dedicated to Prof.~Ian P. Grant and late Prof.~Charlotte Froese Fischer, the creators of the \mrg{original}  {\sc Grasp92} package, 
whose work on atomic theory and computing have been a source of constant inspiration for us. RS and CYC acknowledge support from National Key Research and Development Project of China (No. 2022YFA1602303) and the National Natural Science Foundation of China (No. 12074081 and No. 12104095).  PJ acknowledges support from the Swedish research council under contract 2023-05367. MG acknowledges support from the FWO and F.R.S.-FNRS under the Excellence of Science (EOS) programme (No.~O022818F).

\appendix

\section{Scripts}\label{sec:scripts}
To further test {\sc Graspg}, and to validate the program operation, we provide six scripts.
After installation, the scripts reside in the  \per{ {\tt grasp-master/graspgtest}} subdirectory. The scripts perform calculations 
based on the {\sc Graspg} programs as well
as the corresponding {\sc Grasp2018} programs, making it possible to benchmark the packages against each other with respect to time
and memory usage. The scripts all assume that the  {\sc Grasp2018} and {\sc Graspg} executables are on the path. \per{The directories holding the temporary MPI files, as well as the corresponding {\tt disks} files, are automatically created by the scripts. Note that the user has to modify the setting of the {\tt MaxMemPerProcs} variable for 
the CI calculations
in the {\tt csfg} script directories to reflect the available memory on the local machine.}

\per{
The execution time of the first three scripts are rather short, and the scripts aim
to validate the program operation, and thus the obtained energies for {\sc Graspg} and {\sc Grasp2018} should be very close. The remaining scripts take longer time to execute, and they are mainly though of as giving the user insights how to use the different options to do large scale calculations. Before executing the scripts please make sure that the program {\tt jj2lsj\_2024} has been downloaded from github and is available on the path.}

%In addition,  {\tt disks} files should be available defining the directories for the
%temporary MPI files, see  {\sc Grasp2018} manual \cite{GRASP2018_Man} \mbox{section 6.4}. As an alternative to using {\tt disks} files, the
%user may issue the command
%\begin{Verbatim}[fontsize=\footnotesize]
%>>export MPI_TMP="path to temporary MPI files" 
%\end{Verbatim}
%To adhere to the practice in {\sc Grasp2018} we recommend the former.
\subsection*{Script 1 -- Boron-like Mo from MR-SD VV+CV+CC expansions}
The first script computes the excitation energies for states of the $1s^22s^22p$ odd and $1s^22s2p^2$ even configurations.
The calculations are done by parity based on MR-SD VV+CV+CC CSF expansions to correlation orbitals up to $\{6s,6p,\ldots,6g\}$ (in non-relativistic notation). 
Start by inspecting the file
{\tt run.sh} in the  \per{ {\tt grasp-master/graspgtest/script1}} directory. Then issue the command % { time nohup bash -c ./run.sh ; } > output 2>&1 &
\begin{Verbatim}[fontsize=\footnotesize]
>>nohup ./run.sh >& output &
\end{Verbatim}
\noindent
to put the job in the background. 
The real time for {\sc Graspg} using four processors is \cyc{ 4m\,31s}, which should be compared to   \cyc{ 37m\,46s} using {\sc Grasp2018}.
\subsection*{Script 2 -- Beryllium-like C from SD VV expansions}
The second script computes the excitation energies for states of the $1s^22s^2$, $1s^22p^2$, $1s^22s3s$, $1s^22s3d$ even and 
$1s^22s2p$, $1s^22s3p$, $1s^22p3s$ odd configurations.
Both parities are computed at the same time based on SD VV CSF expansions to correlation orbitals up to $\{9s,9p,\ldots,9g\}$. Start by inspecting the file
{\tt run.sh} in the  \per{ {\tt grasp-master/graspgtest/script2}} directory. Then issue the command
\begin{Verbatim}[fontsize=\footnotesize]
>>nohup ./run.sh >& output &
\end{Verbatim}
For this case the real time for {\sc Graspg} using four processors is \cyc{ 4m\,5s}, which should be compared with  \cyc{ 29m\,20s}  using {\sc Grasp2018}. 
\subsection*{Script 3 -- Beryllium-like C from MR-SD VV+CV+CC expansions}
The third script computes the  the excitation energies for states of the $1s^22s^2$, $1s^22p^2$, $1s^22s3s$, $1s^22s3d$ even and 
$1s^22s2p$, $1s^22s3p$, $1s^22p3s$ odd configurations.
Both parities are computed at the same time based on MR-SD VV+CV+CC CSF expansions to correlation orbitals up to $\{9s,9p,\ldots,9g\}$. Start by inspecting the file
{\tt run.sh} in the  \per{ {\tt grasp-master/graspgtest/script3}} directory. Then issue the command
\begin{Verbatim}[fontsize=\footnotesize]
>>nohup ./run.sh >& output &
\end{Verbatim}
For this case the real time for {\sc Graspg} using sixteen processors is \cyc{ 30m\,6s}, which should be compared with \cyc{ 601m\,49s} using {\sc Grasp2018} with the \per{CSF list based on orbitals ordered by principal quantum numbers (normal list as obtained with {\tt rcsfgenerate} program), or with \cyc{450m\,45s} using {\sc Grasp2018} with the CSF list expanded from the CSFG one. The difference in execution time for the two {\sc Grasp2018} calculations is mainly explained by different number of calls to the Lagrange subroutine depending on 
the order of the orbitals.}
%\cyc{(I did Grasp2018 and Graspg calculations by several times, the order of CSFs play an important role for {\sc Grasp2018} CPU time usage. And there are slight differences between the calculated {\sc Grasp2018} energies by themselves. Does the numerical accuracy account for this kind of differences? Should we discard the script3/csf\_CSFGExp and script3\_ZF/csf\_CSFGExp tests? {\bf The different CPU times reported by GRASP2018-rmcdhf calculations using the original rcsfgenerate-CSF, or that expanded from the corresponding CSFG list generated by rcsfggenerate\_csfg, mainly arise from the SETLAGmpi subroutine, in the form calculation, there were many repeated calculations for potentials.}  )}
\subsection*{Script 3 ZF -- Beryllium-like C from MR-SD VV+CV+CC expansions with restricted interaction}
The same as script 3, but now also condensing the CSF list using the {\tt rcsfginteract\_csfg} program, see \cite{GRASP2018_Man} section 3.1, and rearranging the list in a zero-order space and a first-order space
using {\tt rcsfgzerofirst\_csfg} 
to treat some of the interaction perturbatively.
Start by inspecting the file
{\tt run.sh} in the  \per{ {\tt grasp-master/graspgtest/script3\_ZF}} directory. Then issue the command
\begin{Verbatim}[fontsize=\footnotesize]
>>nohup ./run.sh >& output &
\end{Verbatim}
For this case the real time for {\sc Graspg} using sixteen processors is \cyc{ 15m\,51s}, which should be compared with  \cyc{211m\,40s}  using {\sc Grasp2018}.
\subsection*{Script 4 -- Mg-like Fe from MR-SD VV+CV+CC expansions}
This script performs the calculations reported in section \ref{sec:scaling}. Whereas the calculations
based on {\sc Graspg} are fast, the user should be warned that the calculations based on {\sc Grasp2018} take days to finish. Start by inspecting the file
{\tt run.sh} in the  \per{ {\tt grasp-master/graspgtest/script4}} directory. Then issue the command
\begin{Verbatim}[fontsize=\footnotesize]
>>nohup ./run.sh >& output &
\end{Verbatim}
For this case the real time for {\sc Graspg} using twenty processors is \cyc{ 466m\,21s}, which should be compared with  \cyc{ 8803m\,44s} using {\sc Grasp2018}.
\subsection*{Script 5 -- Mg-like Fe from MR-SD VV+CV+CC expansions}
The last script copies the radial wave functions, the CSFG lists and the labeling orbital lists from the {\sc Graspg} calculations in script 4. The
script in the {\tt 01\_block} directory uses the {\tt rci\_block\_csfg\_mpi} program to generate the Hamiltonian matrix based on 40 processors. After a re-distribution
of the {\tt rci.res} files using the {\tt rdistHmatrix\_csfg} program, {\tt rci\_block\_csfg\_mpi} diagonalizes the Hamiltonian matrix in re-run mode based on 20 processors. The script in the 
{\tt 02\_RAC\_LBr} directory uses the {\tt rmixaccumulate\_csfg} program to condense a CSF list from a fast CI calculation based on the Dirac-Coulomb Hamiltonian. The
reduced list is then used for a CI calculations based on the Dirac-Coulomb-Breit Hamiltonian with and without limitations on the Breit interaction.
Finally, the script in the {\tt 03\_RAC\_Ext\_LBr} directory  uses the {\tt rmixaccumulate\_csfg} program to condense a CSF list from a fast CI calculation based on the Dirac-Coulomb Hamiltonian. The condensed list is \cyc{ then extended} by one orbital layer and the resulting CSF list is then used for a CI calculations based on the Dirac-Coulomb-Breit Hamiltonian with limitations on the Breit interaction.
\end{document}